\documentclass[journal=jpccck,manuscript=article]{achemso}

\usepackage[T1]{fontenc}
\usepackage{float}
\usepackage{graphicx}
\usepackage{amsmath,amssymb}
\usepackage[version=4]{mhchem}
\usepackage{booktabs}
\usepackage{multirow}
\usepackage{enumitem}
\usepackage{siunitx}
\usepackage{xcolor}
\usepackage{float}
\usepackage[normalem]{ulem}  
\usepackage{orcidlink}
\DeclareUnicodeCharacter{2081}{$_1$}

\author{Vivek Chowdhury\,\orcidlink{0009-0006-6180-8398}}
\altaffiliation{Vivek Chowdhury and Tarvir Anjum Aditto contributed equally to this work.}
\affiliation[BUET]
{Department of Electrical and Electronic Engineering, Bangladesh University of Engineering and Technology, Dhaka, Dhaka-1205, Bangladesh}

\author{Tarvir Anjum Aditto\,\orcidlink{0009-0007-6430-4600}}
\altaffiliation{Vivek Chowdhury and Tarvir Anjum Aditto contributed equally to this work.}
\affiliation[BUET]
{Department of Electrical and Electronic Engineering, Bangladesh University of Engineering and Technology, Dhaka, Dhaka-1205, Bangladesh}

\author{Md. Samrat\,\orcidlink{0009-0002-2736-9441}}
\affiliation[BUET]
{Department of Electrical and Electronic Engineering, Bangladesh University of Engineering and Technology, Dhaka, Dhaka-1205, Bangladesh}

\author{Hafiz Imtiaz\,\orcidlink{0000-0002-2042-5941}}
\email{hafizimtiaz@eee.buet.ac.bd}
\affiliation[BUET]
{Department of Electrical and Electronic Engineering, Bangladesh University of Engineering and Technology, Dhaka, Dhaka-1205, Bangladesh}

\author{Ahmed Zubair\,\orcidlink{0000-0002-1833-2244}}
\email{ahmedzubair@eee.buet.ac.bd}
\affiliation[BUET]
{Department of Electrical and Electronic Engineering, Bangladesh University of Engineering and Technology, Dhaka, Dhaka-1205, Bangladesh}

\title{Machine learning based prediction of optical properties in two-dimensional Mo-W-S-Se-Te transition-metal dichalcogenide alloys through physics-informed sampling}

\abbreviations{
TMD, transition-metal dichalcogenide;
DFT, density functional theory;
PBE, Perdew--Burke--Ernzerhof;
GGA, generalized-gradient approximation;
PFN, prior-data fitted network;
TabPFN, tabular prior-fitted network;
MAE, mean absolute error
}


\begin{document}



\begin{tocentry}

\centering
\includegraphics[width=8.3cm,height=4.5cm]{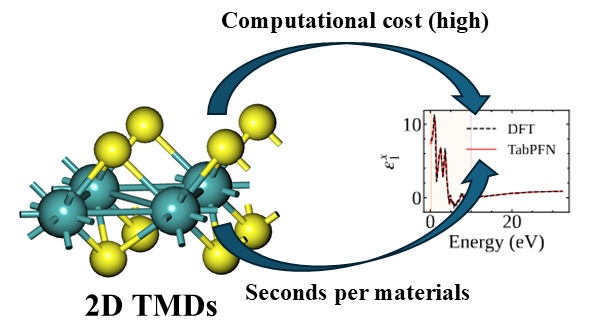}

\end{tocentry}

\begin{abstract}
Two-dimensional transition-metal dichalcogenide (TMD) alloys provide a compositionally tunable platform for controlling the optical and electronic properties. However, systematic prediction of their dielectric response across multicomponent alloy spaces remains challenging owing to the combinatorial cost of first-principles calculations. In this work, we combined \textit{ab initio} optical-property calculations with a tabular foundation-model regression to predict the real and imaginary components of the frequency-dependent dielectric function for Mo--W--S--Se--Te TMD alloys. A dataset of 99 alloy structures spanning binary, ternary, quaternary, and quinary compositions was generated using density functional theory (DFT). The resulting polarization-dependent dielectric spectra were used to train a tabular prior-fitted network (TabPFN) and evaluated against the conventional Extra Trees and XGBoost models. To accommodate the in-context capacity limit of TabPFN, we introduced a non-uniform, physics-informed energy subsampling strategy that concentrates sampling in the optically active region above the band gap, where interband absorption is strongest. Trained solely on quaternary alloys, our TabPFN reconstructed the dielectric spectra of held-out quaternary compositions with an $R^2 > 0.98$ and a mean absolute error below $0.10$ for all four dielectric components, outperforming both baselines while requiring no gradient-based training or hyperparameter tuning. Our model further predicted derived optical quantities, including refractive index, extinction coefficient, and absorption coefficient. Additionally, our model generalized in a zero-shot manner to binary, ternary, and quinary alloys absent from the training set, with quinary predictions achieving an $R^2 > 0.97$. We argue that our proposed approach provides a data-driven route for reconstructing optical spectra from limited first-principles data and offers physical insight into composition-dependent optical responses in low-dimensional TMD alloys.
\end{abstract}


\section{Introduction}


Two-dimensional (2D) materials are interesting candidates for future electronic~\cite{wang2024critical}, optoelectronic~\cite{baugher2014optoelectronic, ciarrocchi2022excitonic}, photonic~\cite{zhang2014electrically}, and valleytronic devices~\cite{ciarrocchi2022excitonic}. Similar to charge-based electronics and spin-based spintronics, the valley degree of freedom of 2D materials can act as information carriers, unveiling the application of valley filters and valves~\cite{PhysRevB.105.L081115}, optical switches~\cite{PhysRevApplied.19.034056}, magnetic switches, and nonvolatile memory devices~\cite{pal2023quantum}. Among different families of 2D materials, transition metal dichalcogenides (TMDs) are especially attractive because of their thickness-dependent band structures, strong spin-orbit coupling, pronounced excitonic effects, and compatibility with van der Waals heterostructures. TMDs are commonly represented by the chemical formula MX$_2$, where M denotes a transition metal atom, such as Mo or W; and X denotes a chalcogen atom, such as S, Se, or Te. Typical semiconducting TMDs include MoS$_2$, MoSe$_2$, WS$_2$, and WSe$_2$, which have been widely studied for nanoscale electronics and optoelectronics~\cite{liu2019valleytronics,huang2022enhanced,jiang2022flexible}. These materials exhibit tunable electronic and optical properties as their thickness is reduced from bulk to few-layer and monolayer limits, making them suitable for field-effect transistors, photodetectors, light-emitting devices, photovoltaic devices, sensors, optical modulators, flexible electronics, and quantum/valleytronic platforms~\cite{jiang2022flexible,li2024photodetector,rani2024advancements}.


Moreover, recent optical-constant measurements have shown that multilayer TMDs can possess high refractive indices, optical anisotropy, and hyperbolic optical responses, which are important for accurate design of nanophotonic and optoelectronic devices~\cite{munkhbat2022opticalconstants}. Recent studies on the optical response of 2D TMDs have therefore focused not only on photoluminescence, circular dichroism and valley polarization, but also on actively tuning these properties for practical device functions. Strategies, such as strain engineering, alloying, defect modification, electrostatic gating, dielectric engineering, twist-angle control and vertical/lateral heterostructure formation, have been used to modulate band gaps, optical absorption, emission intensity, and carrier lifetime~\cite{ciarrocchi2022excitonic,yu2023progress,huang2022enhanced}. In addition, optical and valleytronic properties can be switched and retained, indicating the potential of TMD heterostructures for memory-type optoelectronic and valleytronic devices~\cite{ye2022nonvolatile}.

Because of these features, 2D TMDs are now being explored for a wide range of optical and optoelectronic applications. Their strong light-matter interaction and atomically thin form factor are useful for ultra thin photodetectors, self-powered photodetectors, image sensors, optical modulators, light-emitting devices, excitonic devices, nanophotonic components, flexible optoelectronics, and photovoltaic systems~\cite{li2024photodetector,rani2024advancements,jiang2022flexible}. Additionally, their coupled spin and valley degrees of freedom provide opportunities for valley-based information processing, quantum emitters, and nonvolatile optical memory devices~\cite{liu2019valleytronics,ciarrocchi2022excitonic,ye2022nonvolatile}. Therefore, systematic investigation and engineering of the optical properties of 2D TMDs remain essential for developing next-generation low-dimensional electronic, photonic, optoelectronic, and valleytronic technologies.

Data-driven methods have recently become central to modern materials science, with machine learning (ML) now routinely applied to predict structural~\cite{structural}, mechanical~\cite{mechanical,mechanical2}, and electronic properties\cite{electronic1, electronic2} across diverse materials systems as a faster alternative to computationally expensive DFT calculations. Traditional ML models, such as decision trees, random forests, extra trees, and multilayer perceptrons (MLPs), have been used due to their simplicity, robustness and ability to model nonlinear structure-property relationships~\cite{aditto2026}. These methods have been successfully employed in numerous contexts, including band gap prediction in TMDs~\cite{bandgap1,bandgap2}, stability classification~\cite{stability}, and property screening in 2D materials more broadly~\cite{propertty_screening,propertty_screening2}. These approaches collectively highlight the versatility of classical ML models for materials informatics applications, even when dataset sizes are limited.

Within this emerging landscape, our previous work~\cite{aditto2026} introduced an Extra Trees-based model capable of predicting conduction and valence band energies at multiple $k$-points for binary and ternary 2D TMD alloys. This approach demonstrated high predictive accuracy, with mean squared errors below 0.001 and correlation coefficients $R^2 > 0.99$, underscoring the potential of ensemble tree algorithms for fast and reliable approximation of full band structures~\cite{aditto2026}. While ML models have seen extensive deployment in modeling band gaps, stability, and other electronic characteristics of 2D materials, optical-property prediction remains an underdeveloped but promising frontier, poised to benefit significantly from the continued expansion of data-driven frameworks.  

An early work established bidirectional mappings between optical spectra and material structure~\cite{optical1}, and subsequent efforts extended this to photonic crystal fibers~\cite{optical2}, and calcium boron-zinc glasses~\cite{optical3}, demonstrating the broad applicability of ensemble and regression models across material classes. More recently, ML has been applied to increasingly complex targets. Accurate predictions of molecular optical properties upon aggregation~\cite{optical6}, synthesis-to-emission mappings for carbon dots~\cite{optical7}, and literature-extracted spectral datasets for thin-film property prediction~\cite{optical8} highlight the expanding scope of the field. Two-dimensional materials have also entered this landscape, with ML models predicting optoelectronic properties and absorption spectra from first-principles data~\cite{optical11}. Barhoumi \textit{et al}.~\cite{optical11} predicted absorption spectra but used photon energy as the sole feature, restricting the model to the prediction of only 1 composition.

To the best of our knowledge, no prior work has predicted the optical properties of TMD alloys, including the real and imaginary parts of the dielectric function, and derived parameters such as, refractive index, extinction and absorption coefficient using a data-driven strategy. To bridge this gap, we generated a dataset of optical properties consisting of 99 alloy samples using \textit{ab initio} calculations and employed a neural-network-based strategy. A persistent challenge in employing neural networks for materials science is that convolutional neural networks and other neural architectures often underperform on tabular data~\cite{cnn_tabular}, which is the predominant format in which material-property datasets are organized. This has significantly limited the adoption of neural-network and deep-learning methods in the field.

Recently, TabPFN~\cite{Hollmann2025} has emerged as a promising solution for small tabular datasets. It is a foundation model that leverages an attention mechanism -- the same mechanism that underpins the success of large language models, and is pretrained on a large and diverse collection of tabular datasets. TabPFN is already used in solar energy meteorology~\cite{solar_met}, turbine noise prediction~\cite{turbine_noise}, and even in medical diagnosis~\cite{medical}. It operates through in-context learning: taking the training data directly as context, learning the relationship between features and target variables, and making predictions in a single forward pass, entirely eliminating the need to fine-tuning or updating model parameters. This renders property prediction substantially faster by removing the overhead of conventional training procedures. However, in-context learning in TabPFN poses a practical limitation: it can accommodate up to 10,000 samples as context. Material-property datasets often contain a large number of entries spanning diverse compositions, all of which carry meaningful information about the relationship between material structure and properties. Restricting the context to an arbitrary subset risks losing critical patterns that drive predictive accuracy. To address this, we propose a physics-informed sampling strategy designed to select a representative and informative subset of training samples that faithfully retains the feature-target relationships present in the full dataset. Using this strategy, we enable TabPFN to effectively model the optical properties of a wide range of materials. We further benchmark TabPFN against well-established classical models (Extra Trees and XGBoost) and demonstrate that TabPFN achieves superior predictive performance with negligible training overhead and reduced data requirements. This is supported by the fact that TabPFN successfully predicts the properties of the materials that were not present in the training/context dataset.

\section{Computational Details}
\subsection{Supercell Construction and Alloying}

To construct the alloy database, a monolayer TMD primitive cell is expanded into a $(4 \times 4 \times 1)$ supercell. Since each primitive cell contains one transition-metal atom and two chalcogen atoms, the resulting supercell contains 16 transition-metal sites and 32 chalcogen sites. The transition-metal sublattice is occupied by W and Mo atoms, while the chalcogen sublattice is occupied by S, Se, and/or Te atoms depending on the alloy family. For each target composition, the required numbers of W, Mo, S, Se, and Te atoms are first determined from the desired composition fractions, and the atoms are then randomly substituted over the available metal and chalcogen sites while maintaining the overall TMD stoichiometry.

The composition of the transition-metal sublattice is defined as
\begin{equation}
x = \frac{N_{\mathrm{Mo}}}{N_{\mathrm{W}}+N_{\mathrm{Mo}}}
= \frac{N_{\mathrm{Mo}}}{16},
\end{equation}
where $N_{\mathrm{Mo}}$ and $N_{\mathrm{W}}$ are the numbers of Mo and W atoms in the supercell, respectively. Thus,
\begin{equation}
1-x = \frac{N_{\mathrm{W}}}{N_{\mathrm{W}}+N_{\mathrm{Mo}}}
= \frac{N_{\mathrm{W}}}{16}.
\end{equation}
Accordingly, the number of metal atoms in the supercell can be written as
\begin{equation}
N_{\mathrm{Mo}} = 16x, \qquad N_{\mathrm{W}} = 16(1-x).
\end{equation}
Because the metal sublattice contains 16 sites, the allowed values of $x$ are discretized in steps of $1/16 = 0.0625$ for ternary, quaternary, and quinary alloys.

For ternary metal-alloyed TMDs, a single chalcogen species $\mathrm{Q}$ is used, where $\mathrm{Q}=\mathrm{S}, \mathrm{Se}, \mathrm{Te}$. The ternary compositions are represented as
\begin{equation}
\mathrm{W}_{1-x}\mathrm{Mo}_{x}\mathrm{Q}_{2}.
\end{equation}
Here, $x$ denotes the Mo fraction on the transition-metal sublattice. The special limiting cases of this equation correspond to the binary compounds: for $x=0$, $\mathrm{W}\mathrm{Q}_{2}$, and for $x=1$, $\mathrm{Mo}\mathrm{Q}_{2}$. Therefore, the binary compounds are naturally included as end members of the ternary alloy formula. For each chalcogen species $\mathrm{Q}$, 17 compositions are generated using the possible metal-site fractions from $x=0$ to $x=1$, giving a total of 51 binary and ternary compositions.

For quaternary alloys, two chalcogen species, $\mathrm{Q}$ and $\mathrm{P}$, are distributed over the 32 chalcogen sites. The general formula can be written as
\begin{equation}
\mathrm{W}_{1-x}\mathrm{Mo}_{x}\mathrm{Q}_{2y}\mathrm{P}_{2(1-y)},
\end{equation}
where
\begin{equation}
y = \frac{N_{\mathrm{Q}}}{N_{\mathrm{Q}}+N_{\mathrm{P}}}
= \frac{N_{\mathrm{Q}}}{32},
\end{equation}
and
\begin{equation}
1-y = \frac{N_{\mathrm{P}}}{N_{\mathrm{Q}}+N_{\mathrm{P}}}
= \frac{N_{\mathrm{P}}}{32}.
\end{equation}
Thus, the number of chalcogen atoms in the supercell is
\begin{equation}
N_{\mathrm{Q}}=32y, \qquad N_{\mathrm{P}}=32(1-y).
\end{equation}
The chalcogen pairs considered in this work are $(\mathrm{Q},\mathrm{P}) = (\mathrm{S},\mathrm{Se})$, $(\mathrm{Se},\mathrm{Te})$, and $(\mathrm{S},\mathrm{Te})$. For the quaternary alloy set, $x$ and $y$ are selected within the ranges $0.1875 \leq x \leq 0.625$ and $0.1875 \leq y \leq 0.8125$. Twelve quaternary compositions are generated for each chalcogen-pair combination, resulting in a total of 36 quaternary alloys.

For quinary alloys, three chalcogen species, S, Se, and Te, are simultaneously distributed over the chalcogen sublattice. The general quinary alloy formula can be written as
\begin{equation}
\mathrm{W}_{1-x}\mathrm{Mo}_{x}\mathrm{S}_{2y}\mathrm{Se}_{2z}\mathrm{Te}_{2(1-y-z)}.
\end{equation}
Here, the chalcogen fractions are determined directly from the atom counts as
\begin{equation}
y = \frac{N_{\mathrm{S}}}{N_{\mathrm{S}}+N_{\mathrm{Se}}+N_{\mathrm{Te}}}
= \frac{N_{\mathrm{S}}}{32},
\end{equation}
\begin{equation}
z = \frac{N_{\mathrm{Se}}}{N_{\mathrm{S}}+N_{\mathrm{Se}}+N_{\mathrm{Te}}}
= \frac{N_{\mathrm{Se}}}{32},
\end{equation}
and
\begin{equation}
1-y-z = \frac{N_{\mathrm{Te}}}{N_{\mathrm{S}}+N_{\mathrm{Se}}+N_{\mathrm{Te}}}
= \frac{N_{\mathrm{Te}}}{32}.
\end{equation}
Therefore,
\begin{equation}
N_{\mathrm{S}}=32y, \qquad 
N_{\mathrm{Se}}=32z, \qquad 
N_{\mathrm{Te}}=32(1-y-z).
\end{equation}
The composition variables must satisfy $0 \leq x \leq 1$, $0 \leq y \leq 1$, $0 \leq z \leq 1$, and $y+z \leq 1$. For the quinary alloy set, $x$, $y$, and $z$ are selected within the desired composition window while preserving the above stoichiometric constraint. In the present dataset, 12 quinary compositions are generated using randomly substituted W/Mo atoms on the metal sublattice and S/Se/Te atoms on the chalcogen sublattice.

Overall, our alloy database consists of 51 binary and ternary compositions, 36 quaternary compositions, and 12 quinary compositions, giving a total of 99 alloy structures. For all alloy families, the random substitution procedure is used to generate representative atomic configurations for the target compositions without explicitly enumerating all possible symmetrically inequivalent arrangements.

\subsection{Density Functional Theory Calculations}
The first-principles calculations based on density functional theory are performed using the Cambridge Serial Total Energy Package (CASTEP)~\cite{clark2005first}. The exchange--correlation interaction is treated within the Perdew--Burke--Ernzerhof generalized-gradient approximation (PBE-GGA)~\cite{ernzerhof1999assessment}. Although hybrid functionals, such as HSE, GW, and PBE+U, can provide improved band-gap values, the PBE-GGA functional is adopted throughout this work because these higher-level methods are computationally more demanding. Norm-conserving pseudopotentials are used to describe the electron--ion interactions for W, Mo, S, Se, and Te. A Gaussian smearing width of 0.05 eV is applied to handle partial electronic occupations, and all calculations are carried out using the non-spin-polarized scheme.

The equilibrium structures are obtained using the Broyden--Fletcher--Goldfarb--Shanno quasi-Newton optimization algorithm~\cite{liu1989limited}. For the self-consistent-field calculations, an energy tolerance of $2 \times 10^{-6}$ eV atom$^{-1}$ is used. During geometry optimization, both the ionic positions and the in-plane lattice vectors are relaxed. The convergence thresholds are set to $1.0 \times 10^{-8}$ eV for electronic self-consistency, $1.0 \times 10^{-3}$ eV \AA$^{-1}$ for atomic forces, 0.5 kbar for residual pressure, and $5.3 \times 10^{-7}$ eV for the ionic total-energy change between consecutive optimization steps. A maximum of 100 ionic optimization steps is allowed. To avoid artificial interactions between periodic images along the out-of-plane direction, a vacuum spacing of at least 18--20 \AA{} is introduced.

Since a larger real-space supercell corresponds to a smaller reciprocal-space Brillouin zone, a Monkhorst--Pack $k$-point mesh of $(2 \times 2 \times 1)$ with no shift is used for Brillouin-zone integration during the self-consistent calculations. This relatively coarse $k$-point mesh provides a balance between computational accuracy and cost for the present supercell calculations. The convergence of the total energy was verified with respect to different selected $k$-point densities (see section A of Appendices).

\subsection{Optical-Property Calculations}

The optical properties of the TMD alloy structures are evaluated using the complex, frequency-dependent dielectric function,
\begin{equation}
\varepsilon^{j}(\omega)=\varepsilon_{1}^{j}(\omega)+i\varepsilon_{2}^{j}(\omega),
\end{equation}
where $\varepsilon_{1}^{j}(\omega)$ and $\varepsilon_{2}^{j}(\omega)$ denote the real and imaginary components of the dielectric function, respectively. The superscript $j$ represents the polarization direction. In this work, the optical response is calculated for both the in-plane $x$ direction and the out-of-plane $z$ direction.

The imaginary part of the dielectric function, $\varepsilon_{2}^{j}(\omega)$, describes optical absorption arising from inter-band transitions between occupied valence-band states and unoccupied conduction-band states. It is given by
\begin{equation}
\varepsilon_{2}^{j}(\omega)=
\frac{2\pi e^{2}}{\varepsilon_{0}\Omega}
\sum_{\mathbf{k},v,c}
w_{\mathbf{k}}
\left|
\left\langle
\psi_{c\mathbf{k}}
\left|
\hat{e}^{j}\cdot\mathbf{r}
\right|
\psi_{v\mathbf{k}}
\right\rangle
\right|^{2}
\delta
\left(
E_{c\mathbf{k}}-E_{v\mathbf{k}}-\hbar\omega
\right),
\end{equation}
where $\Omega$ is the unit-cell volume, $e$ is the electronic charge, $\varepsilon_{0}$ is the vacuum permittivity, $\omega$ is the angular frequency of the incident photon, and $w_{\mathbf{k}}$ is the weight of the $k$-point. The term $\hat{e}^{j}$ denotes the unit polarization vector along direction $j$. Here, $\psi_{v\mathbf{k}}$ and $\psi_{c\mathbf{k}}$ are the valence- and conduction-band wavefunctions, respectively, while $E_{v\mathbf{k}}$ and $E_{c\mathbf{k}}$ are their corresponding eigenvalues. The real part of the dielectric function, $\varepsilon_{1}^{j}(\omega)$, was obtained from $\varepsilon_{2}^{j}(\omega)$ through the Kramers--Kronig transformation:
\begin{equation}
\varepsilon_{1}^{j}(\omega)=
1+\frac{2}{\pi}
\mathcal{P}
\int_{0}^{\infty}
\frac{\omega'\varepsilon_{2}^{j}(\omega')}
{\omega'^{2}-\omega^{2}}
d\omega',
\end{equation}
where $\mathcal{P}$ indicates the Cauchy principal value of the integral.

\begin{figure}[H]
    \centering
    \includegraphics[width=0.9\textwidth]{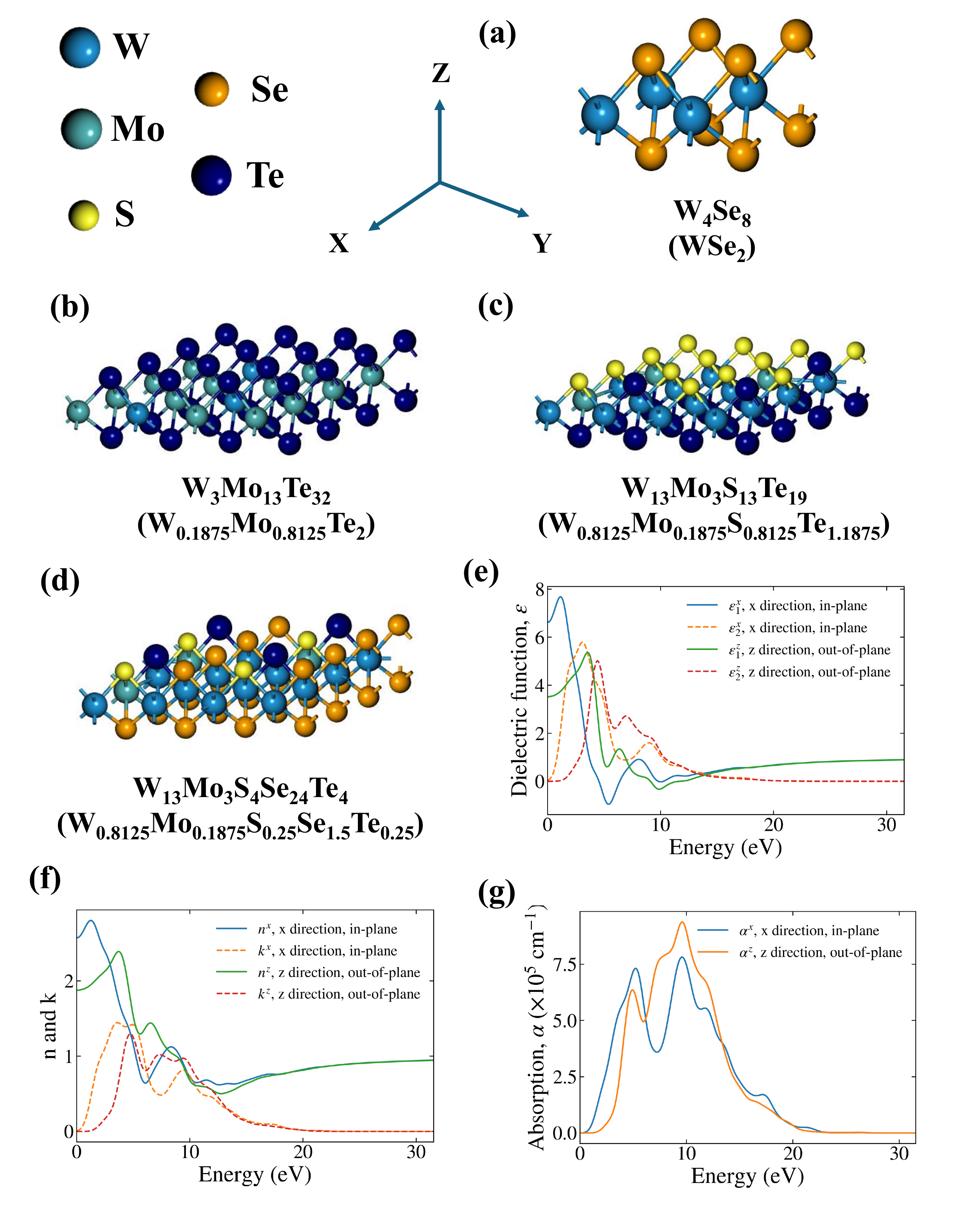}
    \caption{ 
    (a) Binary $\mathrm{WSe_2}$ represented by the $\mathrm{W_4Se_8}$ supercell, 
    (b) metal-alloyed ternary $\mathrm{W_3Mo_{13}Te_{32}}$
    $(\mathrm{W_{0.1875}Mo_{0.8125}Te_2})$, 
    (c) quaternary $\mathrm{W_{13}Mo_3S_{13}Te_{19}}$
    $(\mathrm{W_{0.8125}Mo_{0.1875}S_{0.8125}Te_{1.1875}})$, and 
    (d) quinary $\mathrm{W_{13}Mo_3S_4Se_{24}Te_4}$
    $(\mathrm{W_{0.8125}Mo_{0.1875}S_{0.25}Se_{1.5}Te_{0.25}})$ alloy structures. 
    The color legend identifies W, Mo, S, Se, and Te atoms, and the coordinate axes indicate 
    the in-plane and out-of-plane directions. 
    (e) Real and imaginary components of the dielectric function, 
    $\varepsilon_1$ and $\varepsilon_2$, 
    (f) $n$ and $k$, 
    (g) absorption coefficient $\alpha$, along the in-plane ($x$) and out-of-plane ($z$) polarization directions of $\mathrm{W_3Mo_{13}Te_{32}}$
    $(\mathrm{W_{0.1875}Mo_{0.8125}Te_2})$ from \textit{ab initio} calculations.
    }
    \label{fig:representative_structures_optical_properties}
\end{figure}

For each of the 99 alloy configurations, the calculated optical dataset consists of photon energy-dependent values of $\varepsilon_{1}$ and $\varepsilon_{2}$ for the selected polarization directions. The refractive index, $n^{j}(\omega)$, and extinction coefficient, $k^{j}(\omega)$, are subsequently derived from the dielectric function as
\begin{equation}
n^{j}(\omega)=
\left[
\frac{
\sqrt{\left(\varepsilon_{1}^{j}(\omega)\right)^{2}
+\left(\varepsilon_{2}^{j}(\omega)\right)^{2}}
+\varepsilon_{1}^{j}(\omega)
}{2}
\right]^{1/2},
\end{equation}
and
\begin{equation}
k^{j}(\omega)=
\left[
\frac{
\sqrt{\left(\varepsilon_{1}^{j}(\omega)\right)^{2}
+\left(\varepsilon_{2}^{j}(\omega)\right)^{2}}
-\varepsilon_{1}^{j}(\omega)
}{2}
\right]^{1/2}.
\end{equation}
Using the extinction coefficient, the absorption coefficient is calculated as
\begin{equation}
\alpha^{j}(\omega)=
\frac{2\omega k^{j}(\omega)}{c},
\end{equation}
where $c$ is the speed of light in vacuum. By substituting the expression for $k^{j}(\omega)$, the absorption coefficient can also be written directly in terms of the real and imaginary parts of the dielectric function:
\begin{equation}
\alpha^{j}(\omega)=
\frac{\sqrt{2}\omega}{c}
\left[
\sqrt{\left(\varepsilon_{1}^{j}(\omega)\right)^{2}
+\left(\varepsilon_{2}^{j}(\omega)\right)^{2}}
-\varepsilon_{1}^{j}(\omega)
\right]^{1/2}.
\end{equation}
Representative atomic configurations of the selected TMD systems are shown in 
\autoref{fig:representative_structures_optical_properties}a--d. The binary $\mathrm{WSe_2}$ structure, represented by the $\mathrm{W_4Se_8}$ supercell, shows the characteristic layered arrangement of TMDs
(\autoref{fig:representative_structures_optical_properties}a). The metal-alloyed ternary structure $\mathrm{W_3Mo_{13}Te_{32}}$ 
$(\mathrm{W_{0.1875}Mo_{0.8125}Te_2})$ is formed by substituting W and Mo atoms on the transition-metal sublattice, with Te atoms occupying the chalcogen sites (\autoref{fig:representative_structures_optical_properties}b). 

In comparison, the quaternary $\mathrm{W_{13}Mo_3S_{13}Te_{19}}$$(\mathrm{W_{0.8125}Mo_{0.1875}S_{0.8125}Te_{1.1875}})$ and quinary 
$\mathrm{W_{13}Mo_3S_4Se_{24}Te_4}$ 
$(\mathrm{W_{0.8125}Mo_{0.1875}S_{0.25}Se_{1.5}Te_{0.25}})$ structures 
represent more compositionally complex alloy configurations with mixed metal and chalcogen sublattices 
(\autoref{fig:representative_structures_optical_properties}c,d). The color legend identified W, Mo, S, Se, and Te atoms, while the coordinate axes indicated the in-plane and out-of-plane directions. The optical-property results from \textit{ab initio} calculations shown in 
\autoref{fig:representative_structures_optical_properties}e--g correspond specifically to the ternary alloy $\mathrm{W_3Mo_{13}Te_{32}}$ 
$(\mathrm{W_{0.1875}Mo_{0.8125}Te_2})$. We also calculated the \textit{ab initio} optical property of binary, quaternary, and quinary alloys (see section B of Appendices). The real and imaginary components of the 
dielectric function, $\varepsilon_1$ and $\varepsilon_2$, show a clear dependence on polarization direction 
(\autoref{fig:representative_structures_optical_properties}e). The in-plane ($x$) and out-of-plane ($z$) responses exhibit different peak positions and intensities, indicating anisotropic electronic transitions in the alloy. The refractive index $n$ and extinction coefficient $k$ also vary strongly with photon energy and polarization direction 
(\autoref{fig:representative_structures_optical_properties}f), confirming the direction-dependent light propagation and attenuation behavior of 
$\mathrm{W_3Mo_{13}Te_{32}}$. The absorption coefficient $\alpha$ increases rapidly in the low-energy region and exhibits strong absorption in the visible-to-ultraviolet energy range 
(\autoref{fig:representative_structures_optical_properties}g). These results demonstrate that $\mathrm{W_3Mo_{13}Te_{32}}$ possesses pronounced polarization-dependent optical behavior.

\subsection{Tabular Prior-Fitted Networks}
The optical properties of the Mo--W--S--Se--Te alloy system are predicted using the TabPFN v2~\cite{Hollmann2025} -- a transformer-based tabular foundation model that performs regression via in-context learning without any task-specific gradient-based training. 

\subsubsection{Prior-Data Fitted Networks and the Meta-Learning Objective}
TabPFN belongs to the class of prior-data fitted networks~\cite{Muller2022}, in which a neural network is trained not on a single fixed dataset but on a distribution of datasets $\mathcal{P}(\mathcal{D})$, referred to as \emph{the prior}. During meta-training, the model receives a set of labeled examples $\{(\mathbf{x}_i, y_i)\}_{i=1}^{N}$ drawn from a dataset $\mathcal{D}$ sampled from $\mathcal{P}(\mathcal{D})$, together with an unlabeled query point $\mathbf{x}^{*}$, and is trained to predict $y^{*}$ by minimizing the expected negative log-likelihood
\begin{equation}
    \mathcal{L}(\theta)
    =
    \mathbb{E}_{\mathcal{D} \sim \mathcal{P}(\mathcal{D})}
    \,
    \mathbb{E}_{(\mathbf{x}^{*}, y^{*}) \sim \mathcal{D}}
    \!\left[
        -\log\, p_{\theta}\!\left(
            y^{*} \mid \mathbf{x}^{*},\, \mathcal{D}_{\mathrm{train}}
        \right)
    \right],
    \label{eq:pfn_loss}
\end{equation}
where $\mathcal{D}_{\mathrm{train}}$ is the labeled context and $\theta$ denotes the model parameters. We note that this objective is optimized over the distribution of datasets rather than over any single task. As a result, the trained model approximates Bayesian inference over the prior $\mathcal{P}(\mathcal{D})$: given a new, real-world dataset at inference time, the model computes a posterior predictive distribution without performing any additional gradient updates~\cite{statistical_pfn}.

Unlike other large-scale neural networks, PFNs are trained exclusively on synthetic data and rely on real-world data only as conditioning input~\cite{future_pfn}. TabPFN v2 was meta-trained on approximately $1.3 \times 10^{8}$ synthetic tabular datasets generated from structural causal models and Bayesian neural networks, deliberately incorporating realistic data artifacts, such as missing values, heterogeneous feature scales, and class imbalance~\cite{Hollmann2025}. By training on such a wide variety of synthetic data, the model essentially develops an intuition for the kinds of patterns that tend to show up in small-to-medium tabular datasets including the way composition and properties tend to relate to each other in alloy systems.

\subsubsection{In-Context Learning for Tabular Regression}
During inference time, TabPFN operates in a purely forward-pass regime, no weight updates or fine-tuning are performed with the target dataset. The entire training set, here comprising the DFT-computed dielectric spectra of the quaternary Mo–W–S–Se–Te alloys, is concatenated with the query points and fed simultaneously into the transformer encoder, whose weights remain fixed as learned during prior meta-training on synthetic datasets. This paradigm, known as in-context learning (ICL), allows the model to condition its predictions directly on the observed data at runtime, without any gradient-based adaptation. In this sense, TabPFN behaves conceptually analogous to a non-parametric estimator, i.e., the predictions are derived by attending to the full training context rather than compressing it into task-specific parameters. Note that TabPFN remains a parametric model in the strict sense, as its inference mechanism is governed by a fixed set of pre-trained weights~\cite{statistical_pfn}.

Formally, for a regression query $\mathbf{x}^{*}$, the model computes
\begin{equation}
    \hat{y}^{*}
    =
    f_{\theta}\!\left(
        \mathbf{x}^{*}
        \;\Big|\;
        \left\{(\mathbf{x}_i, y_i)\right\}_{i=1}^{N}
    \right),
    \label{eq:icl}
\end{equation}
where $f_{\theta}$ is the frozen pretrained transformer and $N$ is the number of in-context training examples. The prediction is produced in a single forward pass with $\mathcal{O}(N^{2})$ attention complexity over the context~\cite{Hollmann2025}.

\begin{figure}[t]
    \centering
    \includegraphics[width=\textwidth]{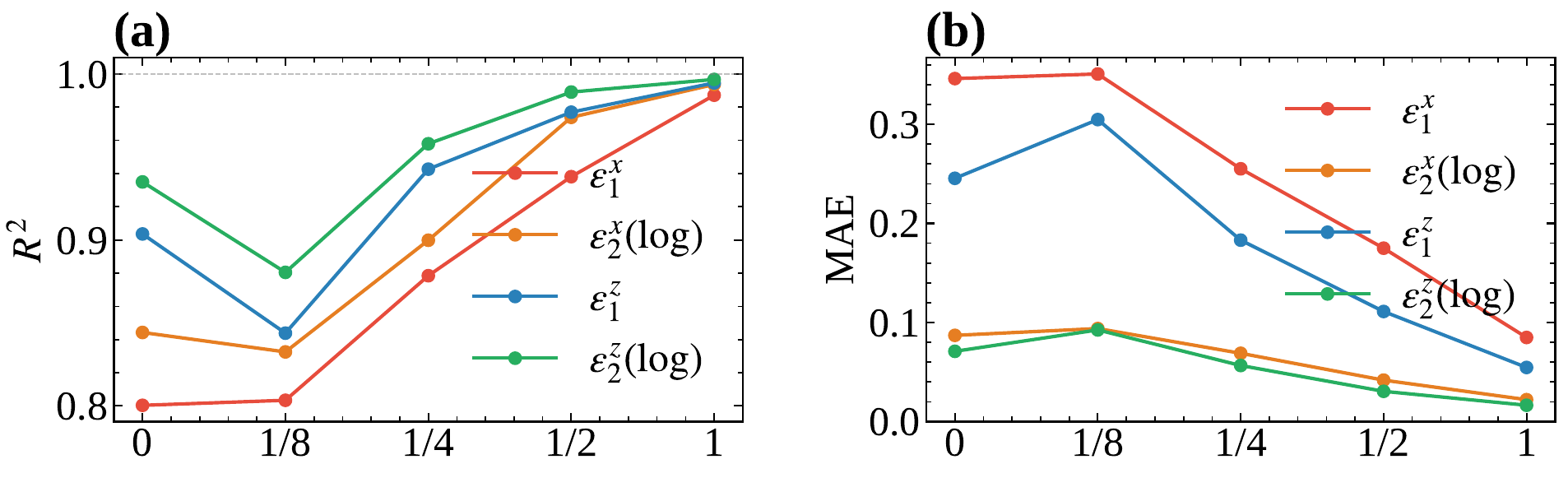}
    \caption{Effect of in-context learning set size on TabPFN prediction 
    performance. (a) $R^2$ and (b) MAE are shown as a function of the 
    fraction of training materials included in the TabPFN context (0, 
    $\frac{1}{8}$, $\frac{1}{4}$, $\frac{1}{2}$, and 1), with smart 
    energy subsampling applied at each fraction (Zero set uses data of only one composition for initializing). All four dielectric 
    components, $\epsilon_1^x$, $\epsilon_2^x$ (log), $\epsilon_1^z$, 
    and $\epsilon_2^z$ (log) , show a consistent and steep improvement 
    beyond $\frac{1}{4}$ of the training materials, converging to 
    $R^2 > 0.98$ and MAE $< 0.1$ at full context, demonstrating the 
    strong sensitivity of TabPFN to in-context sample diversity.}
    \label{fig:tabpfn_icl_scaling}
\end{figure}
Our dataset consists of 36 quaternary alloys, of which 70\% (25 alloys) are used as the training set (for xGBoost and Extra Trees) or in-context set (for TabPFN). The rest of the quaternary alloys, along with binary, ternary, and quinary alloys, are used for testing. \autoref{fig:tabpfn_icl_scaling} illustrates the effect of the training set proportion during ICL on model performance. As the number of alloys included during ICL increases, the model's predictive accuracy improves correspondingly.

\subsubsection{Architecture: Alternating Row and Column Attention}
\begin{figure}[t]
    \centering
    \includegraphics[width=\textwidth]{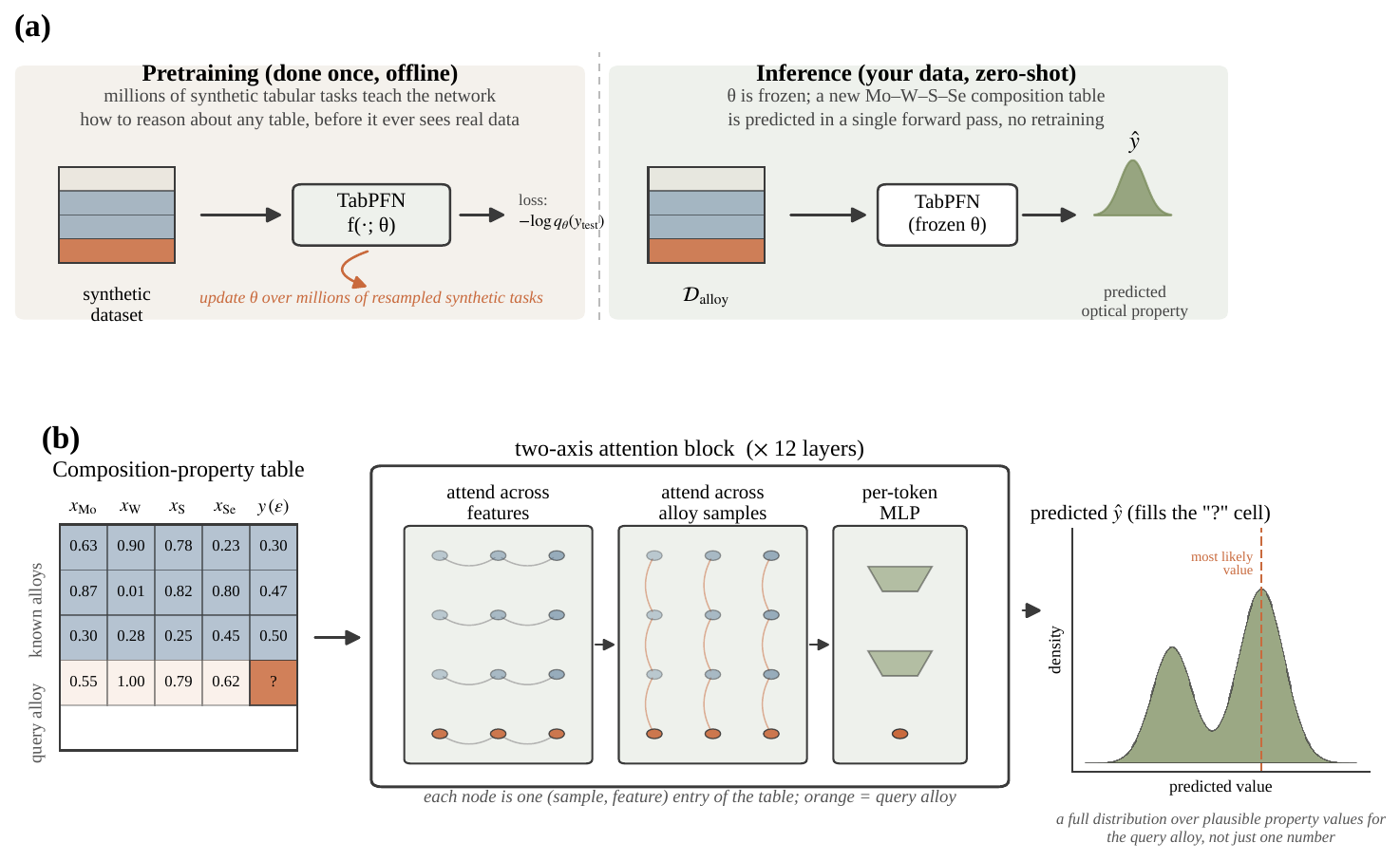}
    \caption{(a) The overview of TabPFN pre-training and inference. (b) Attention mechanism in TabPFN. Image inspired from the original TabPFN article~\cite{Hollmann2025}}
    \label{fig:tabpfn_architecture}
\end{figure}

The transformer encoder in TabPFN v2 departs from the standard token-level self-attention used in natural language processing. Instead, it employs an alternating dual-attention mechanism that operates on a three-dimensional input tensor of shape $(N+1) \times d \times k$, where $N$ is the number of training samples, $d$ is the number of features, and $k$ is the embedding dimension~\cite{Hollmann2025}. \autoref{fig:tabpfn_architecture} illustrates the architecture of TabPFN model.

\begin{description}[leftmargin=0pt, labelindent=0pt]

    \item[Row attention, inter-sample.]
    At each row-attention layer, every sample attends to every other sample across the feature dimension. This allows the model to compare the query point directly against all training examples, effectively computing a learned similarity measure in the embedding space. Compositionally similar alloys, e.g. those sharing similar Mo/W ratios or chalcogen fractions, receive higher attention weights, and their target values contribute proportionally to the query's predicted output.

    \item[Column attention, inter-feature.]
    At each column-attention layer, the model attends across features within a fixed sample. This captures correlations between input descriptors, e.g. between the chalcogen ratio and the energy coordinate, enabling the model to learn physically meaningful feature interactions without explicit feature engineering beyond those provided as inputs.
\end{description}
These two attention types alternate through $L$ transformer blocks ($L = 12$ in the deployed configuration), progressively enriching each token's representation with both cross-sample and cross-feature information. By the final layer, the representation of the query token encodes not only its own feature values but also its relationship to the entire training distribution. This capability proves particularly advantageous for interpolating within the compositional space of the quaternary alloy system.

\subsubsection{Application to Dielectric Spectra Prediction}
The four target quantities, the real and imaginary parts of the dielectric tensor along the in-plane $x$ and out-of-plane $z$ directions, $\varepsilon_{1}^{x}(\omega)$, $\varepsilon_{2}^{x}(\omega)$, $\varepsilon_{1}^{z}(\omega)$, and $\varepsilon_{2}^{z}(\omega)$, were predicted independently. Because the imaginary components were non-negative and exhibited strong right skewness, they were log-transformed prior to the training as
\begin{equation}
    \tilde{\varepsilon}_{2} = \ln\!\left(1 + \varepsilon_{2}\right),
    \label{eq:log_transform}
\end{equation}
and back-transformed via the inverse operation $\varepsilon_{2} = e^{\tilde{\varepsilon}_{2}} - 1$ after prediction.

\subsubsection{Subsampling Strategy}
\begin{figure}[H]
    \centering
    \includegraphics[width=0.9\textwidth]{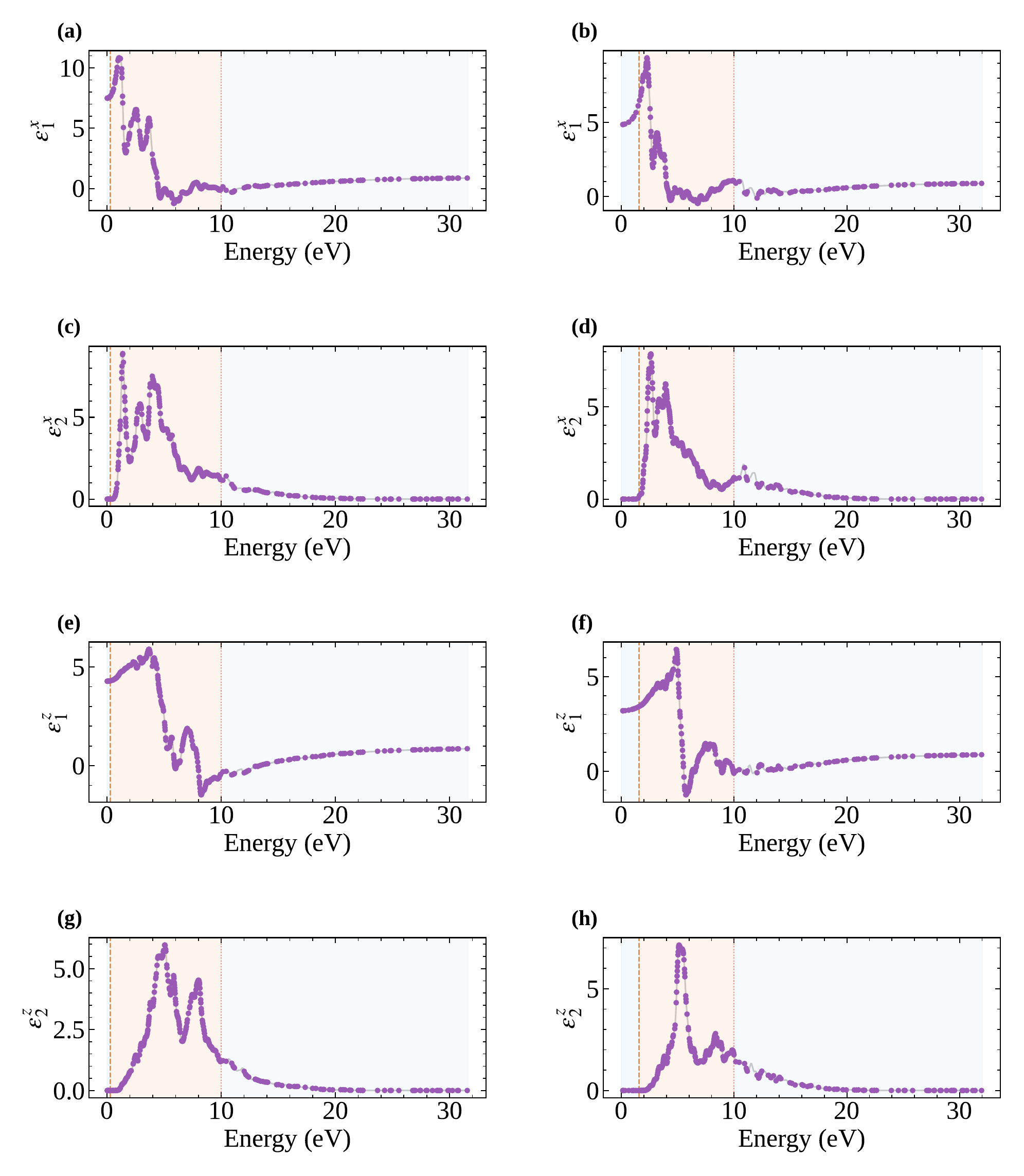}
    \caption{Smart energy subsampling illustrated for two quaternary Mo--W--Se--Te 
    compounds from the training set: Mo$_6$W$_{10}$Se$_{13}$Te$_{19}$ (left, panels a, c, e, g) and Mo$_{5}$W$_{13}$Se$_{6}$Te$_{26}$ (right, panels b, d, f, h). Rows show the real and imaginary parts of the in-plane ($\epsilon_1^x$, $\epsilon_2^x$) and out-of-plane ($\epsilon_1^z$, $\epsilon_2^z$) dielectric functions. The full DFT spectrum (grey, 2000 pts) is overlaid with smart-subsampled points (purple, 330 pts) used as TabPFN training context. The orange dashed and red dotted vertical lines mark $E_g$ and the boundary $\max(10\ \mathrm{eV},\ 3E_g)$, with shaded regions denoting sub-gap (blue, 10 pts), active absorption (orange, 250 pts), and high-energy tail (blue, 70 pts) zones.}
    \label{fig:subsampling}
\end{figure}


The training/context dataset comprised $25$ quaternary alloy compositions, each sampled at approximately 2,000 energy points over the range $0.01$--$33\,\mathrm{eV}$, yielding approximately $52,000$ rows in total. Since the TabPFN's in-context capacity is bounded by practical memory constraints, with ${\leq}\,10,000$ training rows in the standard configuration, we propose a non-uniform energy subsampling strategy that allocates denser sampling above the band gap energy $E_g$, where interband absorption is concentrated, and sparser sampling in the sub-gap and high-energy tail regions (illustrated in \autoref{fig:subsampling}). The details of the approach is presented in the following:

\begin{enumerate}[label=(\roman*), leftmargin=*, itemsep=4pt]

    \item \textbf{Sub-gap region} ($E < E_{g}$): 10 points, sampled uniformly. The dielectric response is slowly varying in this regime, with $\varepsilon_{2} \approx 0$, so sparse coverage is sufficient.

    \item \textbf{Optically active region} ($E_{g} \leq E \leq \max\{10\,\mathrm{eV},\; 3E_{g}\}$): 250 points, sampled uniformly. This window encompasses the absorption edge, the primary excitonic peaks, and secondary interband transitions. The lower bound of $10\,\mathrm{eV}$ was imposed as an absolute floor to ensure adequate coverage for low-band-gap compositions, $E_{g} < 1.67\,\mathrm{eV}$, whose absorption peaks extend beyond $3E_{g}$.

    \item \textbf{High-energy tail} ($E > \max\{10\,\mathrm{eV},\; 3E_{g}\}$): 70 points, sampled uniformly. The dielectric function varies slowly and approaches unity asymptotically in this regime, making dense sampling unnecessary.

\end{enumerate}
\begin{table}[t]
    \centering
    \caption{Non-uniform energy subsampling budget per material (330 points total).}
    \label{tab:sampling_budget}
    \begin{tabular}{llcc}
        \hline
        \textbf{Region} & \textbf{Energy range} & \textbf{Points} & \textbf{Fraction} \\
        \hline
        Sub-gap & $E < E_g$ & 10 & 3.0\% \\
        Optically active & $E_g \leq E \leq \max\{10\ \mathrm{eV},\ 3E_g\}$ & 250 & 75.8\% \\
        High-energy tail & $E > \max\{10\ \mathrm{eV},\ 3E_g\}$ & 70 & 21.2\% \\
        \hline
        \textbf{Total} & & \textbf{330} & \textbf{100\%} \\
        \hline
    \end{tabular}
\end{table}
This allocation, 330 points per material (as summarized in~\autoref{tab:sampling_budget}) $\times$ $25$ training materials $\approx$ $8250$ rows, fitted comfortably within TabPFN's context limit while preserving full compositional-space coverage, a critical requirement for accurate interpolation across the Mo-W-S-Se-Te alloy space. The model was evaluated on the full 2,000-point spectra of the held-out test materials, demonstrating that the subsampled context was sufficient for accurate spectral reconstruction at arbitrary energy coordinates. The input feature vector for each material--energy row is summarized in~\autoref{tab:features}. All features were standardized to zero mean and unit variance using parameters fitted exclusively on the training set and applied without modification to the test set and to the quinary generalization dataset.

\begin{table}[t]
\centering
\caption{Input features used for TabPFN regression of the dielectric-tensor components of Mo--W--S--Se--Te alloys.}
\label{tab:features}
\begin{tabular}{llp{6.5cm}}
\toprule
\textbf{Group} & \textbf{Feature} & \textbf{Description} \\
\midrule
\multirow{5}{*}{Composition}
    & $x_{\mathrm{Mo}}$ & Mo atomic fraction \\
    & $x_{\mathrm{W}}$  & W atomic fraction  \\
    & $x_{\mathrm{S}}$  & S atomic fraction  \\
    & $x_{\mathrm{Se}}$ & Se atomic fraction \\
    & $x_{\mathrm{Te}}$ & Te atomic fraction \\
\midrule
Energy & $E$ & Photon energy (eV) \\
\bottomrule
\end{tabular}
\end{table}

\subsection{Evaluation Metrics}
To evaluate model performance, we employed two complementary metrics: the coefficient of determination ($R^2$) and the mean absolute error (MAE), defined as:
\begin{equation}
    R^2 = 1 - \frac{\sum_{i}(y_i - \hat{y}_i)^2}{\sum_{i}(y_i - \bar{y})^2}, \qquad \text{MAE} = \frac{1}{n}\sum_{i=1}^{n}|y_i - \hat{y}_i|
\end{equation}
where $y_i$, $\hat{y}_i$, and $\bar{y}$ denote the true, predicted, and mean target values, respectively. $R^2$ score ranges from 0 to 1, where values closer to 1 indicate strong predictive performance, and values closer to 0 reflect poor generalization. MAE was expressed in the same units as the target variable, with lower values indicating better accuracy; a value of 0 denotes perfect agreement. Together, $R^2$ captures global goodness of fit, while MAE quantifies typical point-wise deviation, providing a comprehensive assessment of model performance.

\section{Results and Discussion}
\subsection{Optical Spectra Prediction}

\begin{figure}[H]
    \centering
    \includegraphics[width=\textwidth]{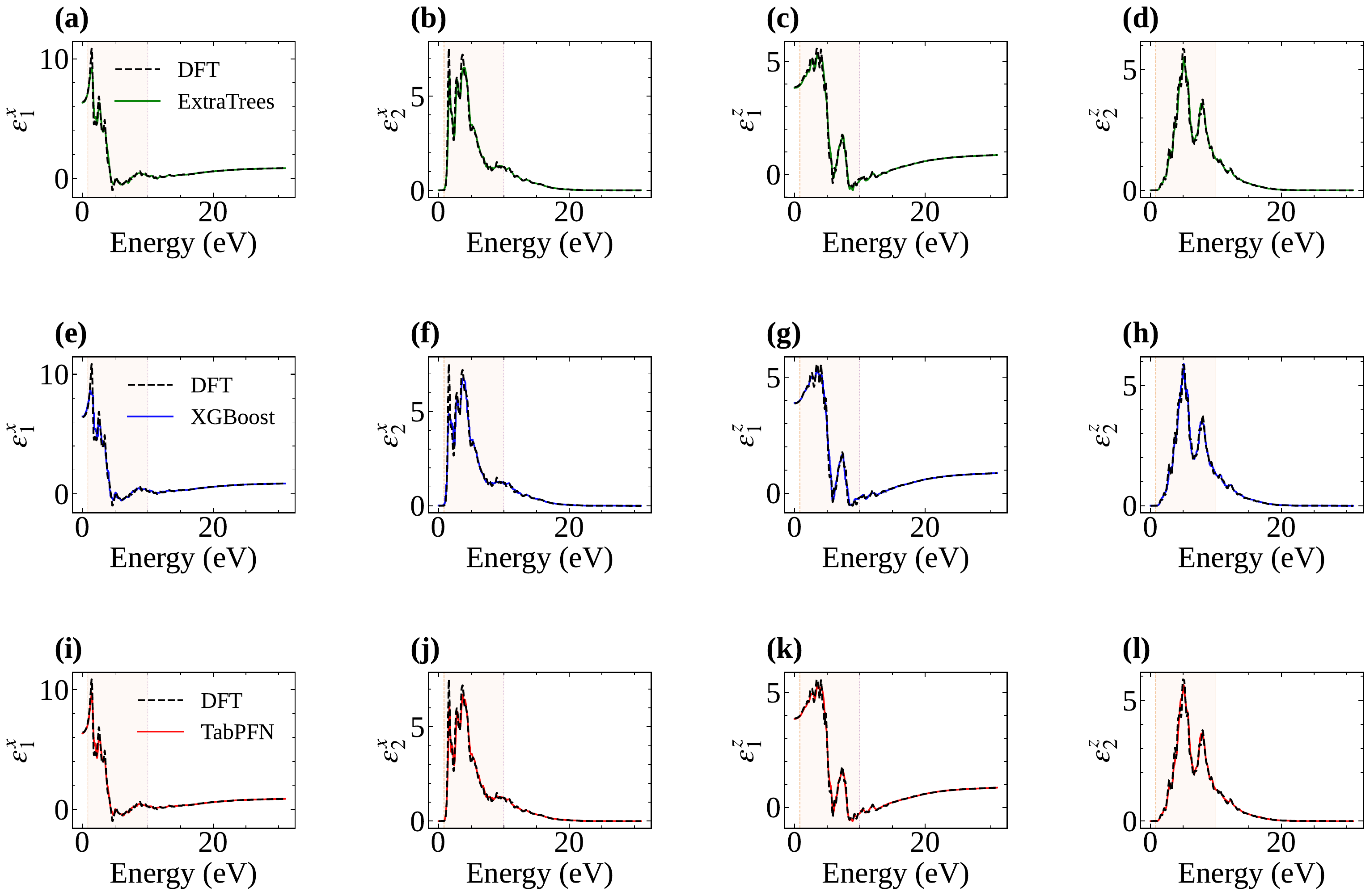}
    \caption{Comparison of DFT-computed (black dashed) and predicted (solid) 
    dielectric spectra for Mo$_{10}$W$_6$S$_{13}$Te$_{19}$ from the test set. 
    Rows correspond to Extra Trees (green, a--d), XGBoost (blue, e--h), and 
    TabPFN (red, i--l). Columns show the real and imaginary parts of the 
    in-plane ($\epsilon_1^x$, $\epsilon_2^x$) and out-of-plane 
    ($\epsilon_1^z$, $\epsilon_2^z$) dielectric functions. The orange shaded 
    region marks the active absorption zone between $E_g$ and 
    $\max(10\ \mathrm{eV},\ 3E_g)$. TabPFN outperforms both XGBoost and Extra Trees in reproducing the DFT computed spectra most faithfully across all components}
    \label{fig:spectrum_comparison}
\end{figure}

We benchmarked TabPFN against Extra Trees \cite{ExtraTrees} and XGBoost \cite{XGBoost} on the held-out quaternary test set comprising seven alloy compositions, each evaluated at approximately 2,000 energy points spanning 0.01--33 eV. \autoref{fig:spectrum_comparison} shows the predicted dielectric spectra for a representative test material (i.e., Mo$_{10}$W$_6$S$_{13}$Te$_{19}$) across all four components -- the real and imaginary parts of the in-plane ($\epsilon_1^x$, $\epsilon_2^x$) and out-of-plane ($\epsilon_1^z$, $\epsilon_2^z$). All three models captured the broad spectral envelope, including the general shape of the absorption onset and the high-energy tail. However, notable differences emerged in the optically active region between $E_g$ and  $\max(10\ \mathrm{eV},\ 3E_g)$, where the dielectric response is governed by interband transitions. Extra Trees and XGBoost approaches reproduced the overall trend but systematically underestimated peak amplitudes and smoothed over the fine spectral features, such as secondary absorption shoulders and post-peak oscillations. This behavior is characteristic of ensemble regression methods, which tend towards the mean of the training distribution and struggle to extrapolate sharp, localized features.~\cite{XGboost_central_bias} TabPFN, in contrast, faithfully reproduced both the position and amplitude of primary and secondary peaks across all four components, with its predicted curves nearly superimposed on the DFT reference throughout the full energy range.

\begin{figure}[H]
    \centering
    \includegraphics[width=\textwidth]{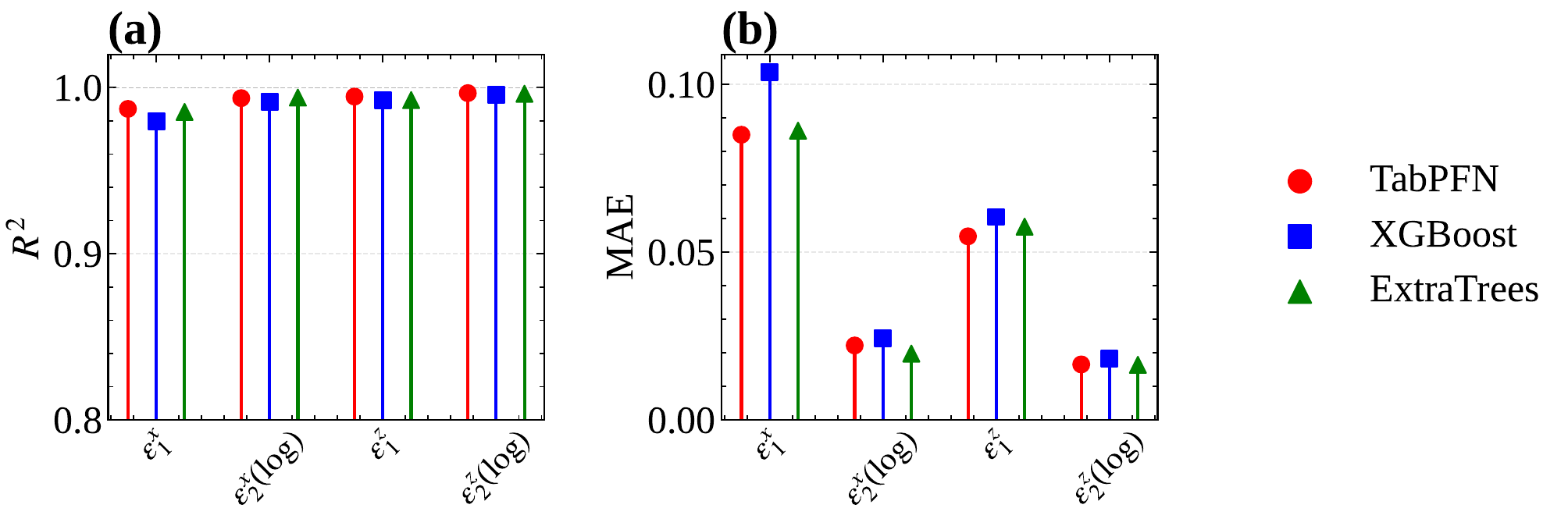}
    \caption{Comparison of $R^2$ (a) and MAE (b) for TabPFN, XGBoost, and 
Extra Trees across the four dielectric function components: 
$\varepsilon_{1}^{x}$, $\varepsilon_{2}^{x}$, $\varepsilon_{1}^{z}$, and $\varepsilon_{2}^{z}$. 
TabPFN achieves the highest $R^2$ and lowest MAE in almost all cases. (MAE for $\varepsilon_{2}^{x}$ and $\varepsilon_{2}^{z}$ is reported in log-transformed units)}
    \label{fig:model_metrics}
\end{figure}

The quantitative performance across the full test set was summarized in \autoref{fig:model_metrics}. TabPFN achieved an $R^2$ score of $> 0.98$ and an MAE of $< 0.10$ for all four dielectric components, outperforming both XGBoost and Extra Trees approaches in almost all cases. Specifically, for $\epsilon_1^x$, TabPFN attains $R^2 = 0.989$ and MAE $= 0.085$, compared to $R^2 = 0.983$ and MAE $= 0.091$ for XGBoost, and $R^2 = 0.986$ and MAE $= 0.082$ for Extra Trees. The imaginary components, $\epsilon_2^x$ (log) and $\epsilon_2^z$ (log), showed a consistent ranking with TabPFN leading in both $R^2$ and MAE scores. The out-of-plane components $\epsilon_1^z$ and $\epsilon_2^z$ were predicted with slightly higher accuracy than their in-plane counterparts across all models. Notably, TabPFN achieved this performance without any hyperparameter tuning, gradient-based training, or task-specific fine-tuning; its inference was completed in a single forward pass over the training context, making it substantially faster in deployment than conventional machine-learning pipelines that required cross-validated model selection.

\begin{figure}[H]
    \centering
    \includegraphics[width=0.9\textwidth]{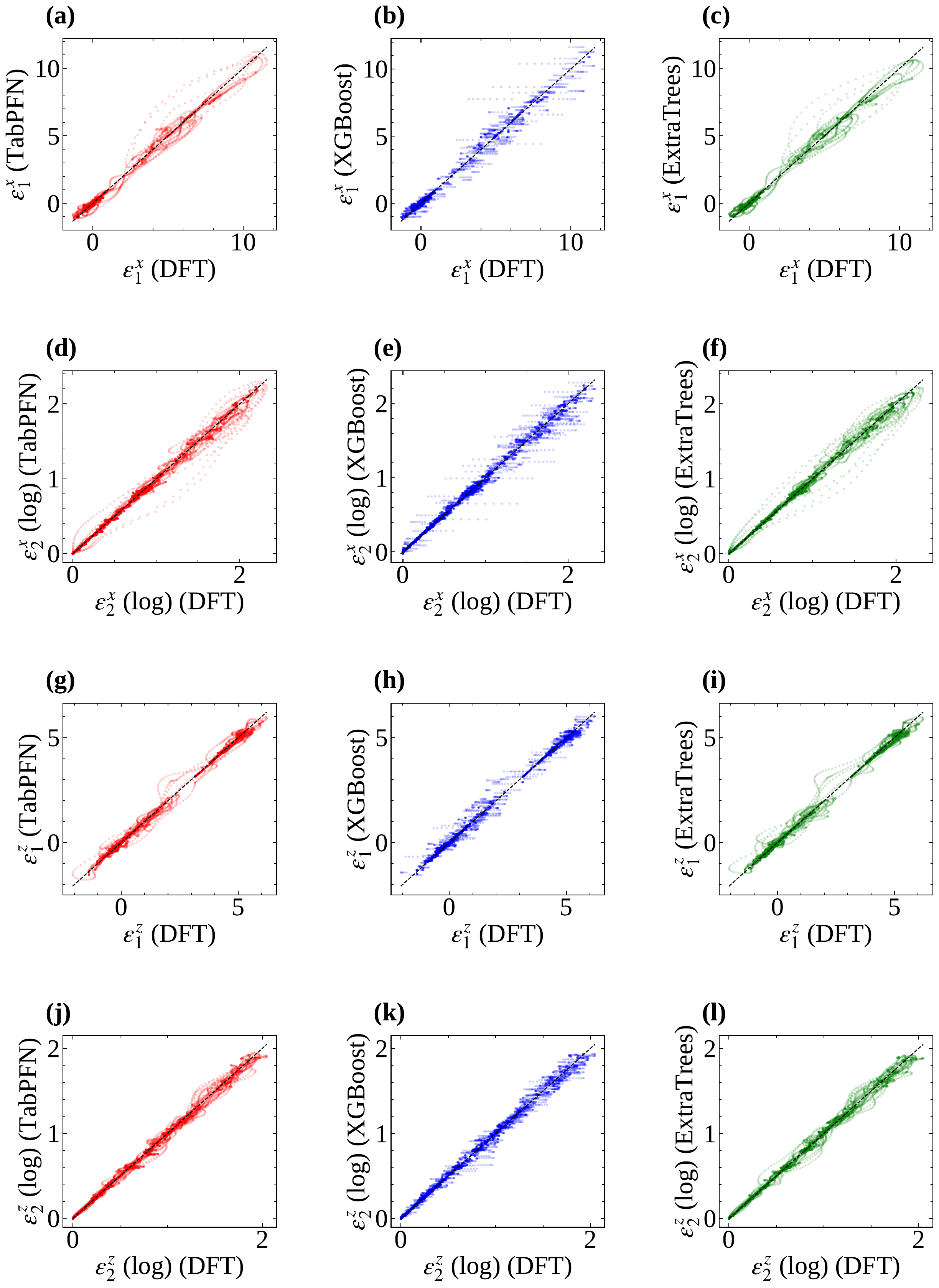}
    \caption{Model-predicted dielectric vs DFT computed dielectric
    function components across the full test set. Rows correspond to 
    $\epsilon_1^x$ (a--c), $\epsilon_2^x$ in log-space (d--f), 
    $\epsilon_1^z$ (g--i), and $\epsilon_2^z$ in log-space (j--l). 
    Columns correspond to TabPFN (red), XGBoost (blue), and Extra Trees 
    (green). The black dashed line denotes perfect agreement.}
    \label{fig:parity_plots}
\end{figure}

The parity plots in \autoref{fig:parity_plots} provide a complementary view of model accuracy across the entire test distribution. In these plots, perfect prediction corresponds to all points lying exactly on the black dashed diagonal $(y=x)$, while increasing scatter away from this line reflects greater prediction error. For TabPFN (red), the predicted values clustered tightly along the ideal parity line across the full dynamic range of each target variable, including the extremes of $\epsilon_1^x$ that reach values above 10. XGBoost (blue) showed slightly broader scatter, particularly at high values of $\epsilon_1^x$ and $\epsilon_1^z$, where predictions were systematically pulled toward the center of the training distribution, a well-known bias of gradient-boosted trees on small datasets.~\cite{XGboost_central_bias} Extra Trees (green) exhibited the widest scatter and the most pronounced regression-to-the-mean behavior, with visible under-prediction at the upper tail of both real components. These deviations are physically consequential: errors in the peak amplitude of $\epsilon_1(\omega)$ propagate directly into the refractive index $n(\omega)$, and errors in $\epsilon_2(\omega)$ affected the extinction coefficient $k(\omega)$ and absorption coefficient $\alpha(\omega)$, all of which are critical quantities for optical device design and material screening.

\begin{figure}[H]
    \centering
    \includegraphics[width=\textwidth]{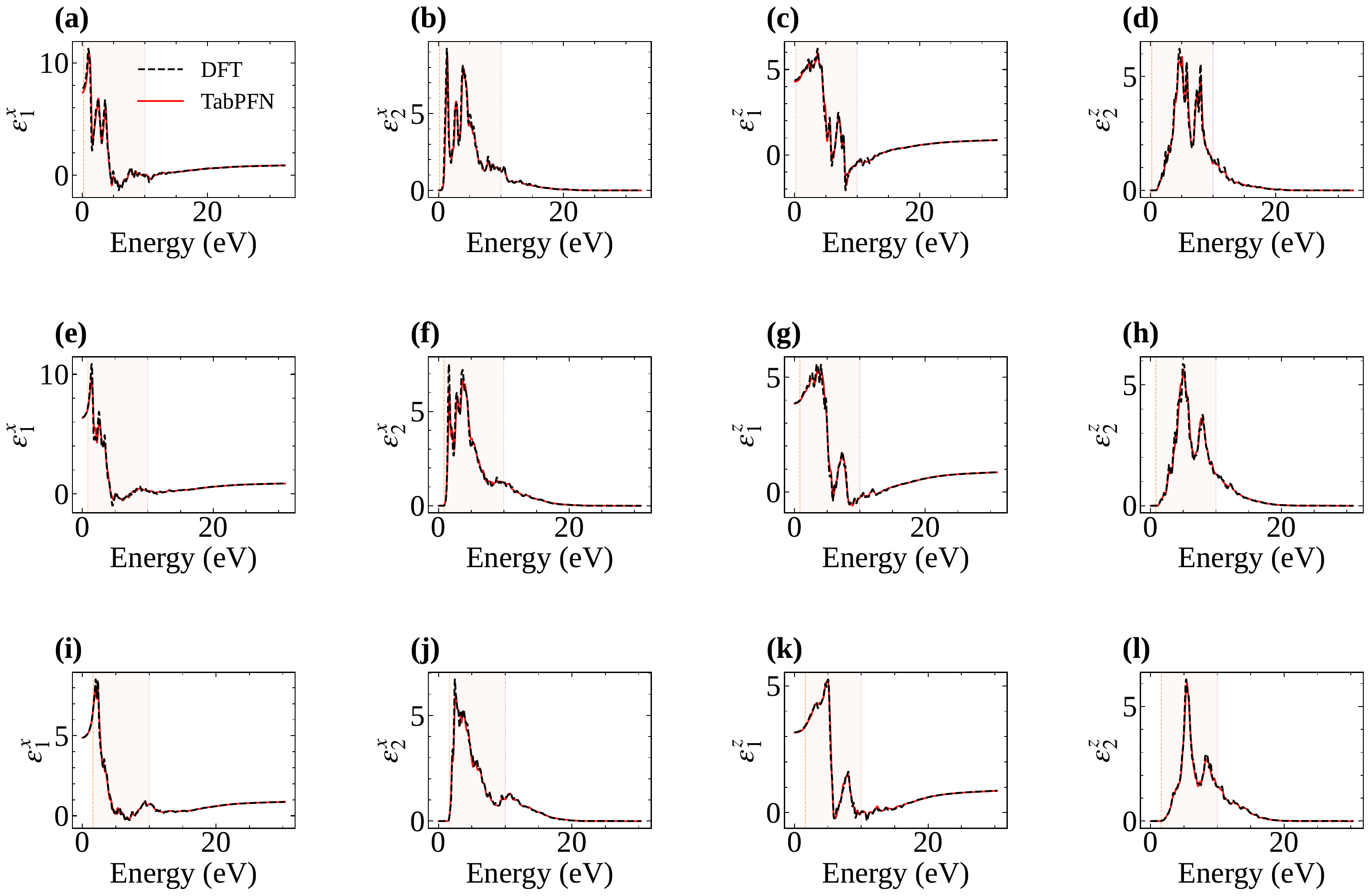}
    \caption{TabPFN-predicted (red solid) vs.\ DFT-computed (black dashed) 
    dielectric spectra for three quaternary compounds from the 
    test set: Mo$_6$W$_{10}$Se$_{6}$Te$_{26}$ (top, a--d), 
    Mo$_{10}$W$_{6}$S$_{13}$Te$_{19}$ (middle, e--h), and 
    Mo$_{6}$W$_{10}$S$_{26}$Te$_{6}$ (bottom, i--l). 
    TabPFN accurately reproduces the DFT 
    spectra across all four components and all three compounds, including 
    fine spectral features near the absorption onset.}
    \label{fig:tabpfn_spectrum_comparison}
\end{figure}

\autoref{fig:tabpfn_spectrum_comparison} shows TabPFN predictions for three additional quaternary alloys from the test set: Mo$_6$W$_{10}$Se$_{6}$Te$_{26}$,  Mo$_{10}$W$_{6}$S$_{13}$Te$_{19}$, and Mo$_6$W$_{10}$S$_{26}$Te$_6$ to further assess spectral fidelity beyond a single test material. These materials spanned a range of Mo/W ratios and chalcogen compositions, and their dielectric spectra differed in both the position of the absorption onset and the number and intensity of interband peaks. Across all three compounds and all four dielectric components, TabPFN predictions were nearly indistinguishable from the DFT reference curves, including the sharp transitions and the fine oscillatory structure in the post-peak region. This consistency across compositionally diverse test materials confirmed that TabPFN is not merely interpolating between close neighbors in the training set but is learning a physically meaningful mapping among composition, electronic-structure descriptors, photon energy, and the dielectric response. The model's ability to simultaneously reproduce both the real and imaginary components further suggested that it captures the underlying relationships of the optical spectra rather than fitting each component independently as an unrelated regression target.

\subsection{Generalization across compositions beyond training data}
\begin{figure}[H]
    \centering
    \includegraphics[width=\textwidth]{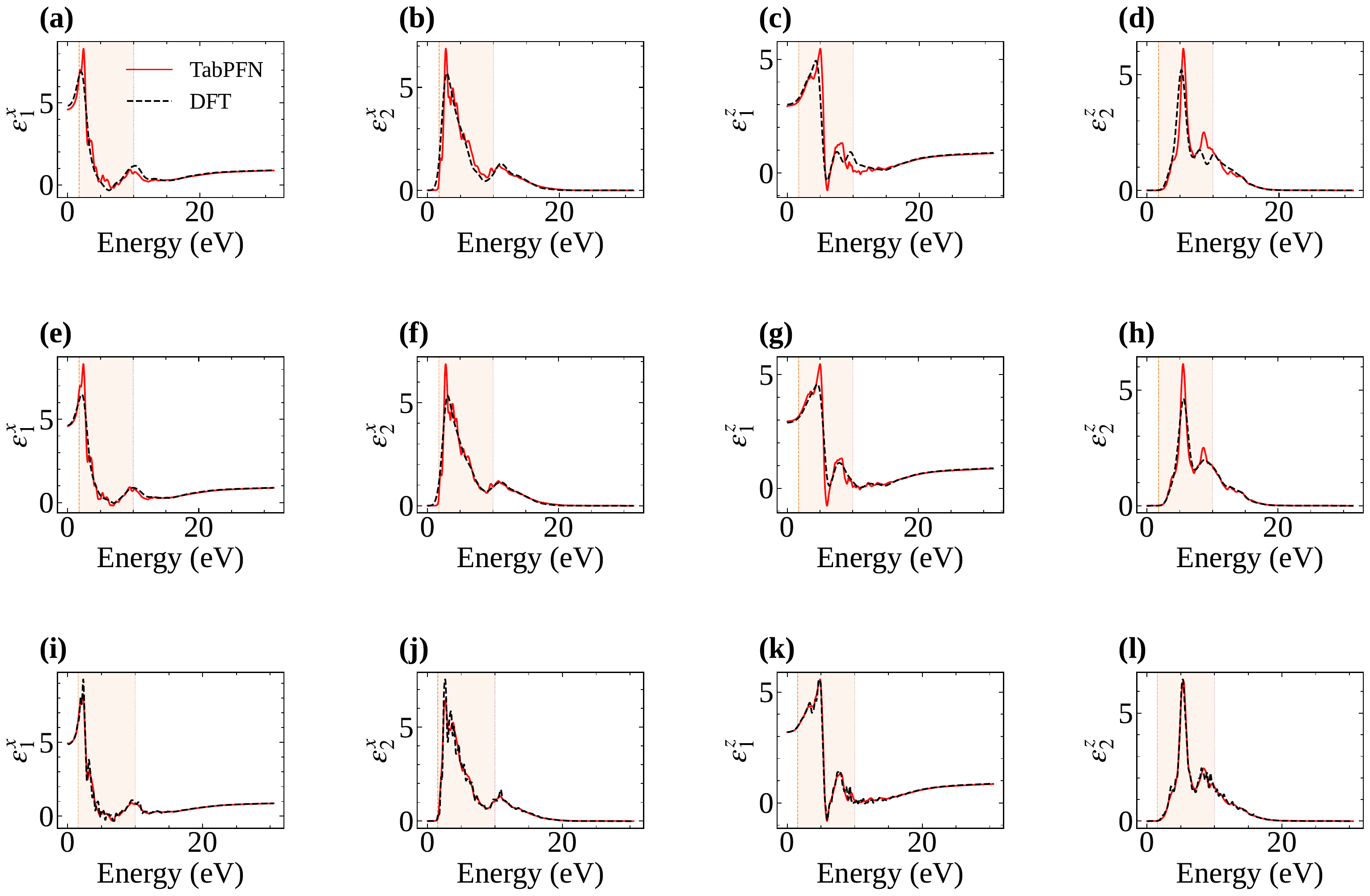}
    \caption{Zero-shot TabPFN predictions (red solid) vs.\ DFT-computed 
    (black dashed) dielectric spectra for binary Mo$_4$S$_8$ (top, a--d), ternary W$_4$Mo$_4$S$_{16}$ (middle, e--h), quinary Mo$_3$W$_{13}$S$_{11}$Se$_{17}$Te$_4$ (bottom, i--l). The model was trained exclusively on quaternary compounds with no binary, ternary or quinary examples in context. TabPFN reproduces the DFT spectra with high fidelity across all components, demonstrating strong out-of-distribution generalization to unseen compositions.}
    \label{fig:quinary_predictions}
\end{figure}

The stringent test of any ML model's generalizability is its ability to make accurate predictions on unseen samples. To this end, the TabPFN's generalizability was tested through prediction of optical properties for binary, ternary, and quinary Mo-W-S-Se-Te alloys, which were absent from the training and in-context learning sets. These materials introduced a new chalcogen species system, expanding the compositional space into a regime that was not seen during training. Predicting optical properties in this setting constituted a zero-shot transfer task: the model must predict beyond its training distribution using only its learned representation of the quaternary alloy space. \autoref{fig:quinary_predictions} shows TabPFN predictions for binary Mo$_4$S$_8$, ternary W$_4$Mo$_4$S$_{16}$, and quinary
Mo$_{3}$W$_{13}$S$_{11}$Se$_{17}$Te$_4$. Despite never having encountered these types of compositions during training, TabPFN reproduced the DFT spectra with high fidelity across all four dielectric components. Though the results were more accurate in the quinary alloy than the binary/ternary alloys, the absorption onset positions, peak amplitudes, and high-energy tails were well-captured, with predicted curves closely tracking the reference data throughout the 0-33~eV energy range. This result was physically significant: the introduction of a third chalcogen species (quinary) or removing chalcogen species (binary/ternary) modifies the band structure, effective masses, and interband transition matrix elements in ways that were not directly represented in the training data, yet the model successfully generalizes to these changes through its learned compositional representations.

Across the three out-of-distribution alloy families, TabPFN's zero-shot generalization improved rather than degraded as compositional complexity increased. As shown in \autoref{tab:alloy_order_generalization}, binary compositions ($n = 6$) and ternary compositions ($n = 45$) show moderate performance, with $R^2$ values clustering between 0.82 and 0.90; and MAE in the 0.15 to 0.25 range across all four dielectric function components. Quinary compositions ($n = 12$), on the other hand, achieved markedly stronger generalization, with $R^2$ consistently above 0.97 and MAE reduced by roughly 60 to 70 percent relative to the binary and ternary cases. This pattern was somewhat counterintuitive given that TabPFN was trained exclusively on quaternary compositions. One might expect prediction quality to degrade monotonically as the target composition departs further from the training distribution in terms of element count. Instead, the out-of-plane components ($\varepsilon_1^{z}$, $\varepsilon_2^{z}$) and the quinary family in particular benefitted from the additional Te-containing compositions that share more structural and electronic similarity with the quaternary training set than the simpler binary and ternary chalcogenides do. The relatively weaker performance on binary and ternary materials likely reflected their narrower compositional space and fewer mixed-element interactions for the model to draw analogies from, even though they were formally closer to quaternary in terms of element count.

Taken together, the zero-shot results demonstrated that TabPFN, trained exclusively on quaternary alloys, learned compositional representations that transfer meaningfully to lower- and higher-order alloy systems. This was a qualitatively stronger form of generalization than interpolation within a fixed composition family. It suggested that the in-context learning framework can serve as a practical tool for screening optical properties across the full Mo-W-S-Se-Te compositional space without requiring explicit DFT calculations for every alloy subfamily. Given that the number of distinct binary, ternary, and quinary compositions grew combinatorially with the number of elements and sub-lattice sites, the ability to predict their properties from a compact quaternary training set represented a substantial reduction in the computational cost of materials exploration. We also evaluated the frequency-dependent optical response (see section C of the Appendices).

To further improve performance on binary and ternary compositions, a small number of binary and ternary examples could be included in the in-context set, rather than relying purely on zero-shot transfer from quaternary training data. This would let TabPFN to draw on compositional analogies closer to the target alloy family, since its in-context learning mechanism is specifically designed to exploit this kind of few-shot conditioning. A practical screening workflow could follow this procedure: generate a small reference dataset for the target alloy family, apply in-context learning to predict optical properties across a broad set of candidate compositions, identify the subset of materials whose predicted properties most closely match the desired target, and then run classical DFT computations only on that narrowed candidate set to obtain validated values. This approach can substantially reduce computational cost by avoiding exhaustive first-principles screening across every possible composition.

\begin{table}[t]
\centering
\caption{Zero-shot generalization of TabPFN across alloy order (no in-context examples for the target alloy family). MAE values are reported in the original linear scale. }
\label{tab:alloy_order_generalization}
\begin{tabular}{lc cc cc cc cc}
\toprule
& & \multicolumn{2}{c}{$\varepsilon_1^{x}$} & \multicolumn{2}{c}{$\varepsilon_2^{x}$} & \multicolumn{2}{c}{$\varepsilon_1^{z}$} & \multicolumn{2}{c}{$\varepsilon_2^{z}$} \\
\cmidrule(lr){3-4} \cmidrule(lr){5-6} \cmidrule(lr){7-8} \cmidrule(lr){9-10}
\textbf{Alloy Order} & \textbf{$n$ materials} & $R^2$ & MAE & $R^2$ & MAE & $R^2$ & MAE & $R^2$ & MAE \\
\midrule
Binary                  & 6  & 0.878 & 0.246 & 0.825 & 0.244 & 0.860 & 0.231 & 0.819 & 0.184 \\
Ternary  & 45 & 0.883 & 0.228 & 0.836 & 0.218 & 0.897 & 0.207 & 0.871 & 0.152 \\
Quinary                 & 12 & 0.986 & 0.082 & 0.970 & 0.095 & 0.992 & 0.063 & 0.988 & 0.061 \\
\bottomrule
\end{tabular}
\end{table}


\section{Conclusion}
In this work, we developed a DFT-informed machine-learning framework for predicting the optical response of two-dimensional Mo--W--S--Se--Te TMD alloys directly from composition. A dataset of 99 binary, ternary, quaternary, and quinary alloy structures was formulated, and the complex dielectric function was computed along the in-plane and out-of-plane polarization directions, from which the refractive index, extinction coefficient, and absorption coefficient were derived. Using a non-uniform energy subsampling strategy to respect its in-context capacity, we proposed to employ TabPFN, which reconstructed the dielectric spectra with high fidelity, achieving an $R^2 > 0.98$ and an MAE $< 0.10$ across all four dielectric components on the held-out quaternary test set. TabPFN consistently surpassed the Extra Trees and XGBoost baselines -- this performance was attained through in-context learning in a single forward pass, without gradient-based training, fine-tuning, or hyperparameter optimization, making the model substantially faster to deploy than conventional pipelines.

The most notable outcome was the model's zero-shot generalization: trained exclusively on quaternary alloys, TabPFN accurately predicted the spectra of binary, ternary, and quinary compositions that were entirely absent from its context, with quinary predictions exceeding $R^2 = 0.97$. Because in-context learning condition predictions (on the full training context rather than compressing it into task-specific parameters), the model learns transferable, physically meaningful compositional representations rather than merely interpolating between neighboring training points. These results suggest that in-context learning with tabular foundation models can serve as a practical, low-cost tool for screening optical properties across the full Mo--W--S--Se--Te compositional space, reserving explicit DFT calculations for a narrowed set of promising candidates. Future work incorporating a small number of lower-order examples into the context, together with additional electronic-structure descriptors, may further improve accuracy on binary and ternary systems, and extend the framework to broader families of two-dimensional materials.

\section*{Supporting Information}
Supporting Information: computational convergence tests, DFT calculated and TabPFN predicted optical properties of the binary, quaternary, and quinary structures.

\begin{acknowledgement}
V.C., T.A.A., M.S., H.I., and A.Z. thank the Research and Innovation Centre for Science and Engineering (RISE), Bangladesh University of Engineering and Technology (BUET), for financial support through the Internal Research Grant (Project ID: 2023-02-022).
\end{acknowledgement}



\bibliography{ref}


\clearpage
\appendix

\renewcommand{\thesection}{S\arabic{section}}
\renewcommand{\thesubsection}{S\arabic{section}.\arabic{subsection}}
\renewcommand{\thefigure}{S\arabic{figure}}
\renewcommand{\thetable}{S\arabic{table}}
\renewcommand{\theequation}{S\arabic{equation}}
\setcounter{section}{0}
\setcounter{figure}{0}
\setcounter{table}{0}
\setcounter{equation}{0}

\section*{Appendices}

\subsection{\textbf{A. Computational Convergence Tests}}

\begin{figure}[H]
    \centering
    \includegraphics[width=0.7\textwidth]{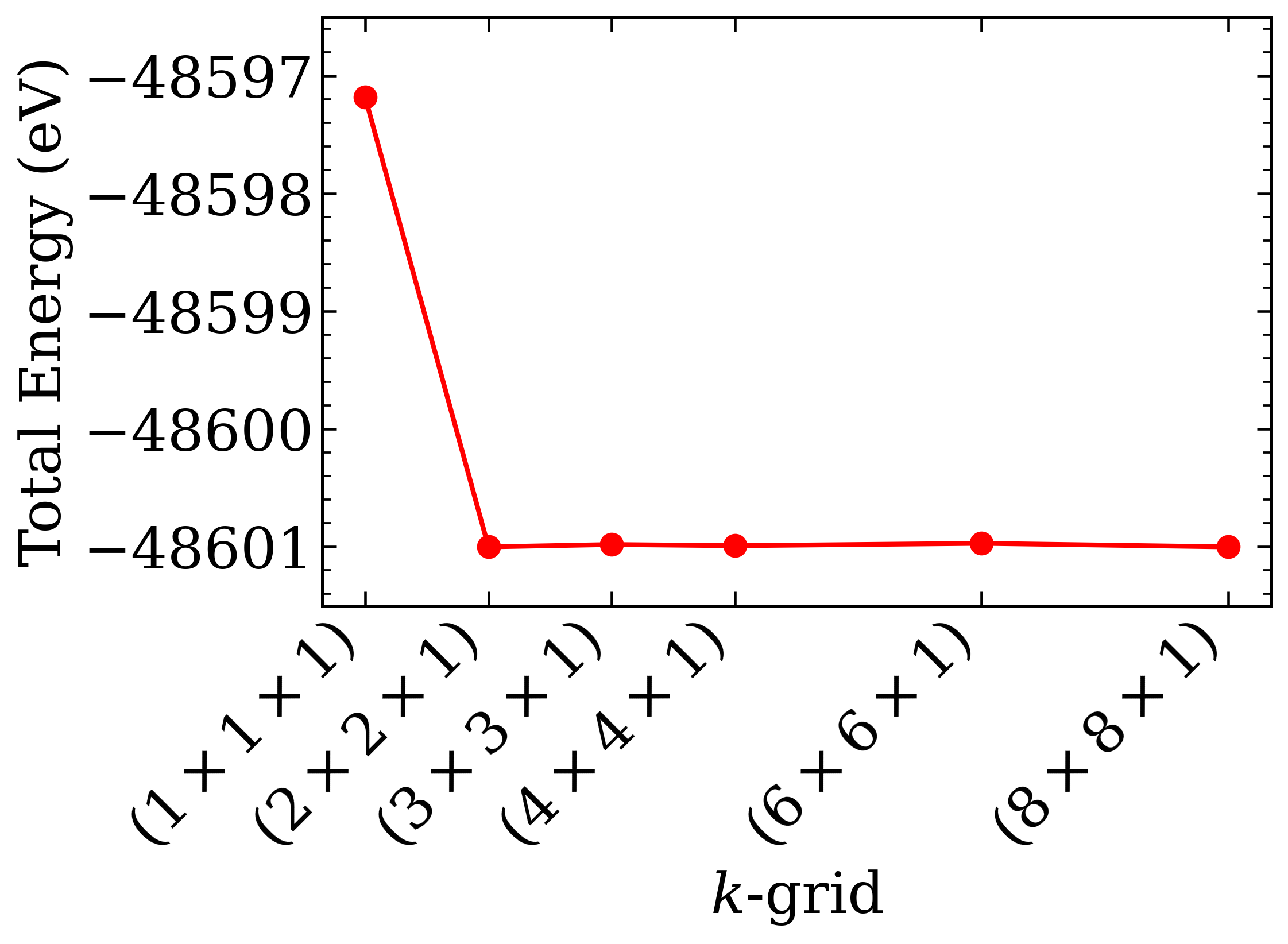}
    \caption{Total energy (eV) variation with respect to different $k$-grid
    sizes of WS$_2$.}
    \label{fig:kgrid_convergence}
\end{figure}

\autoref{fig:kgrid_convergence} depicts a series of Monkhorst--Pack $k$-grids ranging from $(1\times1\times1)$ to
$(8\times8\times1)$ was tested to determine the sampling density required for a
well-converged ground-state energy. The total energy dropped sharply when the
grid was refined from $(1\times1\times1)$ to $(2\times2\times1)$ and remained
essentially constant for all denser grids, indicating that the energy was
converged to within a negligible margin beyond $(2\times2\times1)$. A
$k$-grid of $(2\times2\times1)$ or finer was therefore sufficient to ensure
reliable total-energy values while keeping the computational cost low.

\subsection{\textbf{B. DFT Calculated Optical Properties of the Binary, Quaternary, and Quinary Structures}}

\begin{figure}[H]
    \centering
    \includegraphics[width=\textwidth]{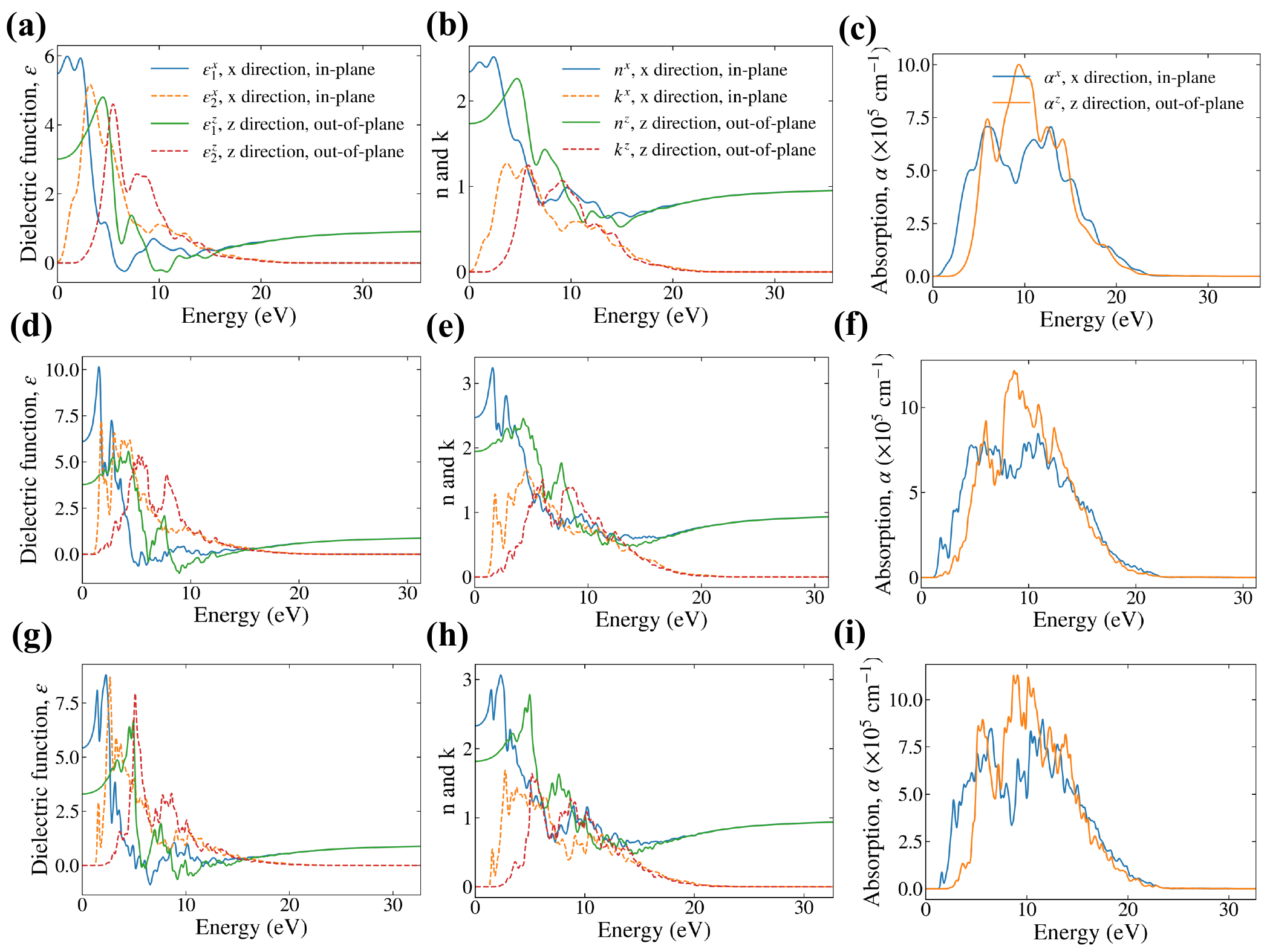}
    \caption{\textit{Ab initio} calculated (a) real and imaginary components of the dielectric function,
    $\varepsilon_1$ and $\varepsilon_2$, (b) refractive index $n$ and
    extinction coefficient $k$, and (c) absorption coefficient $\alpha$ of
    $\mathrm{WSe_2}$ represented by the $\mathrm{W_4Se_8}$ supercell;
    (d) $\varepsilon_1$ and $\varepsilon_2$, (e) $n$ and $k$, and
    (f) $\alpha$ of quaternary $\mathrm{W_{13}Mo_3S_{13}Te_{19}}$
    $(\mathrm{W_{0.8125}Mo_{0.1875}S_{0.8125}Te_{1.1875}})$; and
    (g) $\varepsilon_1$ and $\varepsilon_2$, (h) $n$ and $k$, and
    (i) $\alpha$ of quinary $\mathrm{W_{13}Mo_3S_4Se_{24}Te_4}$
    $(\mathrm{W_{0.8125}Mo_{0.1875}S_{0.25}Se_{1.5}Te_{0.25}})$ alloy. All
    properties are shown along the in-plane ($x$) and out-of-plane ($z$)
    polarization directions as functions of photon energy.}
    \label{supp2}
\end{figure}
This section presents the frequency-dependent DFT calculated optical response of the three
structures studied: the binary $\mathrm{WSe_2}$ reference (represented by the
$\mathrm{W_4Se_8}$ supercell), the quaternary$\mathrm{W_{13}Mo_3S_{13}Te_{19}}$$(\mathrm{W_{0.8125}Mo_{0.1875}S_{0.8125}Te_{1.1875}})$ alloy, and the quinary
$\mathrm{W_{13}Mo_3S_4Se_{24}Te_4}$
$(\mathrm{W_{0.8125}Mo_{0.1875}S_{0.25}Se_{1.5}Te_{0.25}})$ alloy, as shown in
\autoref{supp2}. For each structure we reported the complex dielectric
function, the refractive index and extinction coefficient, and the absorption
coefficient, each resolved into in-plane ($x$) and out-of-plane ($z$)
components to capture the anisotropy of the layered materials.
\autoref{supp2}(a), (d), and (g) present the real and imaginary parts of the
dielectric function, $\varepsilon_1$ and $\varepsilon_2$, for
$\mathrm{WSe_2}$, $\mathrm{W_{13}Mo_3S_{13}Te_{19}}$
    $(\mathrm{W_{0.8125}Mo_{0.1875}S_{0.8125}Te_{1.1875}})$, and
$\mathrm{W_{13}Mo_3S_4Se_{24}Te_4}$
$(\mathrm{W_{0.8125}Mo_{0.1875}S_{0.25}Se_{1.5}Te_{0.25}})$, respectively. The
real part ($\varepsilon_1$) characterized polarization and dispersion, while
the imaginary part ($\varepsilon_2$) reflected optical absorption and interband
transitions. \autoref{supp2}(b), (e), and (h) show the corresponding
refractive index $n$ and extinction coefficient $k$, while \autoref{supp2}(c),
(f), and (i) show the absorption coefficient $\alpha$ for the same three
compounds. Comparing $\mathrm{WSe_2}$, $\mathrm{W_{13}Mo_3S_{13}Te_{19}}$
    $(\mathrm{W_{0.8125}Mo_{0.1875}S_{0.8125}Te_{1.1875}})$, and
$\mathrm{W_{13}Mo_3S_4Se_{24}Te_4}$
$(\mathrm{W_{0.8125}Mo_{0.1875}S_{0.25}Se_{1.5}Te_{0.25}})$ highlighted how
progressive alloying via introducing Mo, S, and finally Te into the
$\mathrm{WSe_2}$ modified the dielectric response, shifted spectral features,
and altered the absorption onset and magnitude.

\subsection{\textbf{C. TabPFN Predicted Optical Properties of the Binary, Quaternary, and Quinary Structures}}
\label{sec:si-tabpfn-optical}
\begin{figure}[H]
    \centering
    \includegraphics[width=\textwidth]{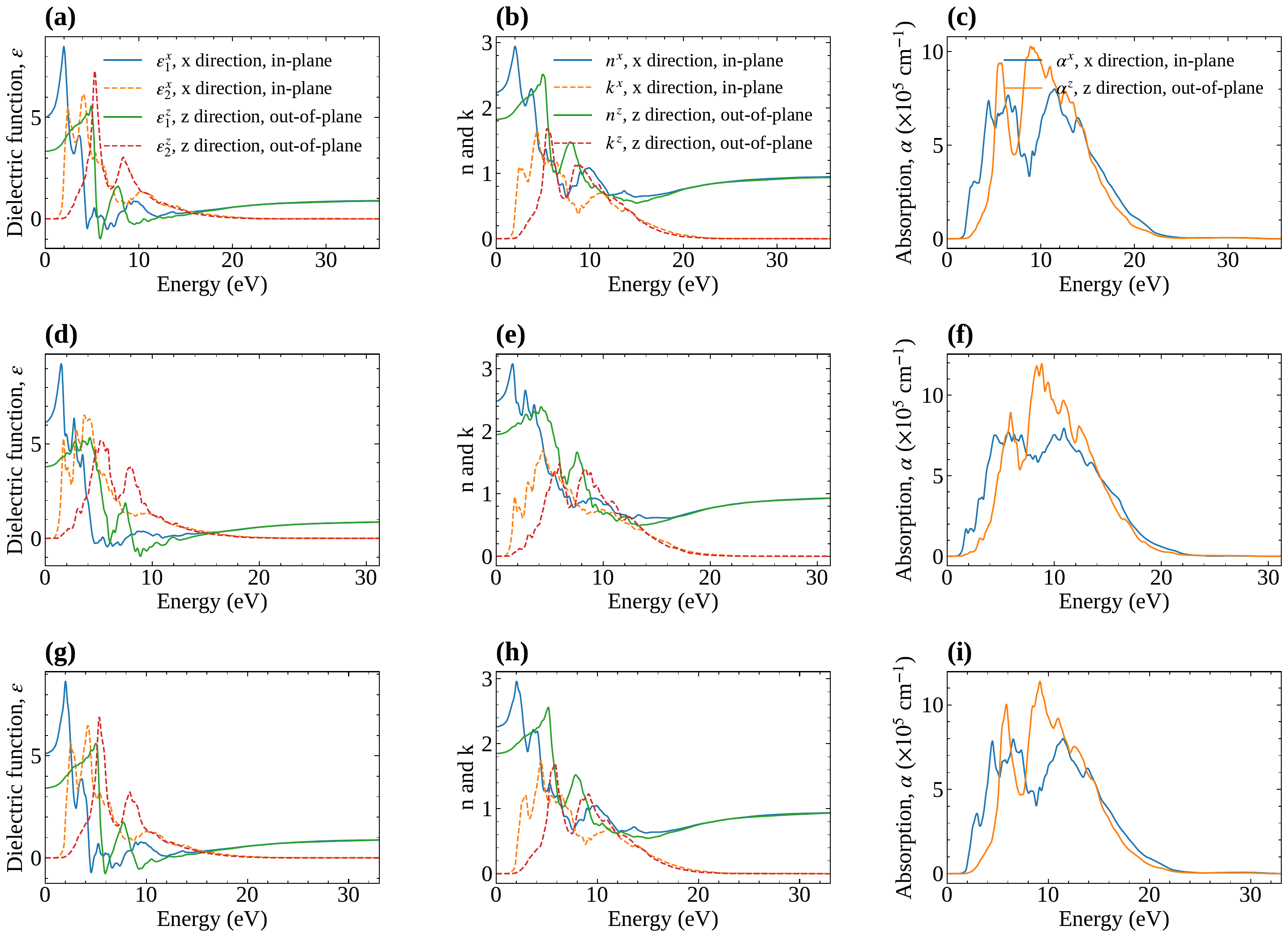}
    \caption{TabPFN predicted (a) real and imaginary components of the dielectric function,
    $\varepsilon_1$ and $\varepsilon_2$, (b) refractive index $n$ and
    extinction coefficient $k$, and (c) absorption coefficient $\alpha$ of
    $\mathrm{WSe_2}$ represented by the $\mathrm{W_4Se_8}$ supercell;
    (d) $\varepsilon_1$ and $\varepsilon_2$, (e) $n$ and $k$, and
    (f) $\alpha$ of quaternary $\mathrm{W_{13}Mo_3S_{13}Te_{19}}$
    $(\mathrm{W_{0.8125}Mo_{0.1875}S_{0.8125}Te_{1.1875}})$; and
    (g) $\varepsilon_1$ and $\varepsilon_2$, (h) $n$ and $k$, and
    (i) $\alpha$ of quinary $\mathrm{W_{13}Mo_3S_4Se_{24}Te_4}$
    $(\mathrm{W_{0.8125}Mo_{0.1875}S_{0.25}Se_{1.5}Te_{0.25}})$ alloy. All
    properties are shown along the in-plane ($x$) and out-of-plane ($z$)
    polarization directions as functions of photon energy.}
    \label{supp3}
\end{figure}
This section presents the frequency-dependent optical response derived from TabPFN predictions for three representative structures: the binary
$\mathrm{WSe_2}$ reference (represented by the $\mathrm{W_4Se_8}$ supercell), the quaternary $\mathrm{W_{13}Mo_3S_{13}Te_{19}}$
($\mathrm{W_{0.8125}Mo_{0.1875}S_{0.8125}Te_{1.1875}}$) alloy, and the quinary $\mathrm{W_{13}Mo_3S_4Se_{24}Te_4}$
($\mathrm{W_{0.8125}Mo_{0.1875}S_{0.25}Se_{1.5}Te_{0.25}}$) alloy, as shown in \autoref{supp3}. Comparison with \autoref{supp2} shows that the predicted response was highly accurate for the quaternary material, whereas the predictions degraded slightly for the binary and quinary materials. This behavior was expected; TabPFN captured the derived properties of quaternary materials with high fidelity because only quaternary alloy data were present in the
in-context set. Zero-shot predictions for the binary and quinary alloys captured the overall spectral trends but often struggle to distinguish closely spaced peaks. Performance could be further improved by including binary, ternary, and quinary alloy data in the in-context learning set.

\end{document}